\documentclass[fleqn,usenatbib]{mnras}

\usepackage{newtxtext,newtxmath}

\usepackage[T1]{fontenc}
\DeclareRobustCommand{\VAN}[3]{#2}
\let\VANthebibliography\thebibliography
\def\thebibliography{\DeclareRobustCommand{\VAN}[3]{##3}\VANthebibliography}

\usepackage{pdflscape}
\usepackage{xurl}
\usepackage{graphicx}	
\usepackage{amsmath}	
\usepackage{xargs}      
\usepackage[pdftex,dvipsnames]{xcolor}
\usepackage[colorinlistoftodos,prependcaption]{todonotes}

\title[CNN Architecture Comparison for Radio Galaxy Classification]{CNN Architecture Comparison for Radio Galaxy Classification}

\author[Burger Becker]{Burger Becker$^{1}$\thanks{Contact e-mail: \href{mailto:adolfburgerbecker@gmail.com}{adolfburgerbecker@gmail.com}}, Mattia Vaccari$^{2}$, Matthew Prescott$^{2}$, Trienko Grobler$^{1}$\\
$^{1}$Computer Science Department, Stellenbosch University, Stellenbosch, South Africa\\
$^{2}$Inter-University Institute for Data Intensive Astronomy (IDIA) and Department of Physics and Astronomy, University of the Western Cape, Robert Sobukwe Road,\\ 7535 Bellville, Cape Town, South Africa\\
}

\date{Accepted XXX. Received YYY; in original form ZZZ}

\pubyear{2020}

\begin{document}
\label{firstpage}
\pagerange{\pageref{firstpage}--\pageref{lastpage}}
\maketitle

\begin{abstract}
The morphological classification of radio sources is important to gain a full understanding of galaxy evolution processes and their relation with local environmental properties. Furthermore, the complex nature of the problem, its appeal for citizen scientists and the large data rates generated by existing and upcoming radio telescopes combine to make the morphological classification of radio sources an ideal test case for the application of machine learning techniques. One approach that has shown great promise recently is Convolutional Neural Networks (CNNs). Literature, however, lacks two major things when it comes to CNNs and radio galaxy morphological classification. Firstly, a proper analysis of whether overfitting occurs when training CNNs to perform radio galaxy morphological classification using a small curated training set is needed. Secondly, a good comparative study regarding the practical applicability of the CNN architectures in literature is required. Both of these shortcomings are addressed in this paper. Multiple performance metrics are used for the latter comparative study, such as inference time, model complexity, computational complexity and mean per class accuracy. As part of this study we also investigate the effect that receptive field, stride length and coverage has on recognition performance. For the sake of completeness, we also investigate the recognition performance gains that we can obtain by employing classification ensembles. A ranking system based upon recognition and computational performance is proposed. {\sc MCRGNet}, {\sc Radio Galaxy Zoo} and {\sc ConvXpress} (novel classifier) are the architectures that best balance computational requirements with recognition performance.
\end{abstract}




\section{Introduction}

Morphological classification is a fundamental aspect of galaxy formation and evolution studies, where the shape of galaxies is intimately connected to the dynamical and physical processes at play. Since the days of Hubble, astronomers have thus been establishing increasingly sophisticated classification schemes to group galaxies in different classes according to their shapes and appearance observed at optical wavelengths \citep{Hubble1926, deVaucouleurs1959, Sandage1961, Elmegreen1987}.

According to our current understanding of galaxy formation and evolution, every massive galaxy is believed to contain a supermassive black hole which undergoes periods of accretion throughout cosmic time to produce an Active Galactic Nucleus (AGN). AGN are often detected in radio surveys via their synchrotron emission produced by accelerated electrons in their cores, lobes and jets, and are then referred to as radio-loud AGN.

Morphologically, \cite{Fanaroff1974} found that radio-loud AGN could be divided into two populations, known as Fanaroff–Riley (FR) types I and II (FRIs and FRIIs), which were found to show a division at approximately $L_{178 {\rm MHz}}= 10^{25}$ W Hz$^{-1}$. Those having bright cores, or "core-brightened" features, and diffuse lobes are labelled as FRIs and those dominated by "edge-brightened" features far from their cores are known as FRIIs. A clear divide in the radio and optical luminosities between the two morphologies was observed in \cite{Owen1994}, indicating that they have formed and evolved in different ways. The different morphological types are thought to be due to different accretion modes. FRIs were believed to be more associated with Low Excitation Radio Galaxies (LERGs); passive galaxies undergoing inefficient accretion of hot gas, with an absence of emission lines in their optical spectra. FRIIs are more associated with High Excitation Radio Galaxies (HERGs); galaxies undergoing rapid and efficient accretion of a cold gas supply, as indicated via the presence of emission lines in their spectra \citep{Hine1979,Laing1994,Best2012,Pracy2016}. While multi-wavelength observations are essential to better pinpoint the centre of radio galaxies and study their physical properties \citep{Prescott2018,Ocran2020,KW2020}, determining radio morphologies is a very useful starting point toward improving our understanding of radio galaxies.

Radio morphologies can also be used to trace the environments of their host galaxies. \cite{Miraghaei2017} found that, at fixed stellar mass and radio luminosity, FRIs are more likely to be found in richer environments than FRIIs. Bent-tailed radio galaxies such as Narrow-Angle Tailed (NAT, \cite{Rudnick1977}) and Wide-Angle Tailed (WAT, \cite{Owen1976,Missaglia2019}) radio galaxies are associated with clusters of galaxies and represent galaxies with radio jets that are interacting with the hot intra-cluster medium (ICM) that resides there. 

New surveys with lower flux limits show that upon closer inspection the FRI/FRII divide becomes less clear, revealing that there is much more overlap in the properties of FRIs and FRIIs than previously thought \citep{Mingo2019}. The FRI/FRII divide is also complicated by radio sources that have hybrid FRI/FRII morphologies, also known as HyMoRS, that have also been found to exist \citep{Gopal2000}. These however are likely to be bent FRII sources viewed at a particular orientation, and have lobes that appear as having different morphologies due to the observer's line of sight \citep{Smith2019,Harwood2020}.

In more recent times, the FR classification scheme has been expanded to include radio sources with compact morphologies. These so-called FR0 sources are believed to be the most abundant radio sources in the local Universe \citep{Baldi2018,Garofalo2019}. Despite being abundant, little is known about their nature. Whilst some are young AGN that will grow to form FRI and FRII sources, a comparison between their number densities and that of extended sources indicates the majority must be older sources that have failed to form extended structures \citep{Sadler2014, Baldi2018}. \cite{Whittam2020} show FR0s are a mixed population of HERG and LERG radio sources.   

Radio sources with more exotic morphologies have also attracted substantial interest recently. These include X-shaped and S-shaped radio galaxies \citep{Cheung2007}, that may represent AGN that have undergone the process of hydrodynamical backflow \citep{Leahy1984} or are the result of a spin-flip from the coalescence of two black holes \citep{Ekers1978}, with the former scenario being the preferred explanation in some of the latest work \citep{Roberts2018,Cotton2020a}.

\subsection{Big Data in Radio Astronomy}

Radio astronomy is currently undergoing a rapid development in observational capabilities which is paving the way for the highly anticipated Square Kilometre Array \citep[SKA]{SKA2015,SKA2019}.

Before the advent of the SKA, its pathfinders and precursors \citep{Norris2013} promise to revolutionize our knowledge of the radio sky. Ongoing surveys such as VLASS \citep{Lacy2020} and EMU \citep{Norris2011} are expected to detect 5 and 70 million radio sources, respectively, greatly exceeding the roughly 2.5 million radio sources known to date. Historically, scientific analysis used catalogues compiled by either individuals or small teams \citep{Fanaroff1974}. However, the increasingly large samples of radio sources detected by modern radio telescopes means that the classification of full catalogues by subject-matter experts is no longer a viable option.

One possible solution is the crowdsourcing of labelling to large groups of volunteers, known as citizen science. The first successful large scale citizen science project in galaxy morphology classification was Galaxy Zoo \citep{Lintott2008}, during which participants were asked to label galaxies observed as part of the Sloan Digital Sky Survey on the basis of their morphology. After initial fears of poor public participation, roughly 100,000 participants made 40 million individual classifications in 175 days. Due to the success of the initial project, Galaxy Zoo grew into the larger Zooniverse\footnote{https://www.zooniverse.org/} project, which serves as an online platform for various crowdsourcing projects. The first citizen science project devoted to radio astronomy was {\sc Radio Galaxy Zoo} \citep{Banfield2015}, which aimed to classify radio sources observed in the FIRST survey based on their morphology and identify them with their infrared counterparts observed in the WISE survey. {\sc Radio Galaxy Zoo} demonstrated citizen scientists can help us to further the scientific exploitation of large radio surveys, in the process creating large samples of visually-inspected labelled sources.

\subsection{Application of Deep Learning to Radio Astronomy}
Another solution would be to make use of machine learning techniques to aid in the classification task. Convolutional Neural Networks (CNNs) are a popular choice for image recognition problems in both academia and industry. A CNN is a special type of neural network that learns which features are important to extract from images. These learned features are then employed to perform classification. {\sc AlexNet} is one of the first CNN architecture to achieve human-level performance on the ImageNet Challenge, which involved classifying 1.2 million images into 1000 different classes \citep{deng2009,Krizhevsky2012}.

CNNs were popularised as an automated means of morphological classification in astronomy during the Galaxy Zoo Challenge~\citep{Willett2013}, hosted on the Kaggle data science platform\footnote{\url{https://www.kaggle.com/c/galaxy-zoo-the-galaxy-challenge}}. \cite{Dieleman2015} developed a CNN for this challenge. The CNN \cite{Dieleman2015} developed obtained the highest recognition performance in this challenge. Both {\sc AlexNet} and the study by \cite{Dieleman2015} inspired {\sc Toothless}, the first CNN developed for the morphological classification of radio galaxies \citep{Aniyan2017}. Since then several new CNNs have been developed for morphological classification in radio astronomy, as shown in Table~\ref{tab:arch_keys}.

To construct a CNN model usually requires a lot of training data. If the dataset that is used for training a CNN is too small overfitting occurs. In most of the studies in Table~\ref{tab:arch_keys} training was done on a small dataset. It is, therefore, imperative to determine whether, in the case of radio galaxy morphological classification, overfitting indeed occurs when a small curated dataset is used for training. Having overfitted is, however, easily correctable by simply using a larger training set. The only real danger occurs when an overfitted model is used to predict the performance of an architecture in a real world setting. In this paper, we conduct an experiment to determine if this issue is in fact something which we should be cognisant of going forward. The experiment we propose will make use of two modified datasets from \citet{Ma2019}.

Moreover, most of the studies in Table~\ref{tab:arch_keys} have a similar overall layout. They first present a novel architecture and then they report on the architecture's recognition performance. Furthermore, most of these studies do not analyse inference time or other factors related to computational cost (with the exception of model complexity often being evaluated through the number of trainable parameters).\footnote{{\sc CLARAN}'s computational cost is reported in \cite{Wu2019}.} Some of these studies also focus on the effects of layer composition \citep{Lukic2018}. Most critically, none have looked at how computational cost impacts recognition performance. Analysis of the relationships between these metrics can further our understanding of existing architectures and help develop best practices for finding new architectures. In this paper we also address this shortcoming, i.e. we use a wide variety of computational related cost metrics to asses whether the architectures in Table~\ref{tab:arch_keys} are capable of real time performance, whilst maintaining a high level of recognition performance.

All the architectures, the software used to analyse and test them and the datasets used in the process are made publicly available.\footnote{\url{https://github.com/BurgerBecker/rg-benchmarker}} 

We start our paper by giving a brief overview of CNNs. In Section~\ref{sec:arch} we list all the architectures we used in our study, which also includes a novel architecture. The datasets we make use of are presented in Section~\ref{sec:data} and is structured according to a simplified 4-class morphological classification system used by \citet{Alhassan2018}: Compact sources, FRI, FRII and Bent-tails. The experimental setup of our study is presented in Section~\ref{sec:experiments}. The results pertaining to the overfitting experiment are presented in Section~\ref{sec:overfit_results}. The results of the computational cost versus recognition performance analysis of the architectures in Table~\ref{tab:arch_keys} are presented in Section~\ref{sec:perf_results} (a pragmatic ranking of the architectures is also established). As part of this performance analysis, we also investigate how other factors like receptive field, stride length, coverage (see Section~\ref{sec:metrics}) impact the recognition performance of the architectures in Table~\ref{tab:arch_keys}. For the sake of completeness, we also investigate whether we can achieve recognition performance gains using ensemble classifiers in Section~\ref{sec:perf_results} (also see Section~\ref{sec:ensemble}). Findings are then summarized according to subject in the conclusion.

\section{Deep Learning Algorithms and Metrics}
A brief overview of CNNs is presented in Section~\ref{sec:cnn}. In Section~\ref{sec:metrics} we briefly discuss the different metrics we made use of to evaluate the experiments we conducted.

\subsection{Deep Convolutional Neural Networks}
\label{sec:cnn}
\subsubsection{Convolutional Neural Networks: A Brief History}
\citet{Hubel1968} found that mammalian visual cortices primarily consist of two types of cells, simple cells (that would activate when straight edges had a certain orientation) and complex cells (with a larger receptive field and lower sensitivity to orientation). This inspired the Neocognitron artificial neural network \citep{Fukushima1980} which combined layers consisting wholly of one of two types of ``cells'' into a hierarchical model to perform handwritten character recognition. One type of cell would apply a convolutional operation to the input, while the other would downsample the input. The weights for the convolutional operation would be learned from input examples. The first deep CNN \citep{LeCun1998} had seven layers and was primarily used for handwritten character recognition. At the time, training such deep models was computationally expensive and time consuming. The widespread advent of Graphics Processing Units (GPU) within desktop computers, however, resulted in more people being able to quickly train deep CNNs \citep{cirecsan2010}. Furthermore, Deep Learning in general and CNNs in specific became the \emph{de facto} standard for image classification after the 2012 ImageNet Challenge was won by a CNN, {\sc AlexNet} \citep{Krizhevsky2012}.

\subsubsection{Overview of Neural Networks}

Neural Networks can be visualized as graph-like structures in which nodes are referred to as neurons. Each edge or connection to another neuron has a weight and a bias term, which represents the strength of their connection. The weights and biases affect the propagation of information through the network.  With the right combination of weights, the network can match groups of similar input, or classes, to a respective output, or class label, reliably. The neurons are arranged in layers, with an input being propagated from the input layer, through intermediate layers (called hidden layers) until it reaches an output layer. Although many different types of neural networks have developed over time, CNNs have become some of the top performing classifiers for image recognition.

CNNs have three main types of layers: Fully connected (or dense) layers, convolutional layers and pooling layers. 

\subsubsection{Fully Connected Layers}
The neurons of a fully connected layer are, as the name implies, connected to all the neurons of the previous layer. The output layer is also a fully connected layer, with as many neurons as the number of classes that have been provided. The neuron with the highest output value determines the classification result, with a higher value meaning greater model confidence in the classification.

\subsubsection{Convolutional Layers}
In convolutional layers, a convolution operation is performed on a small neighbourhood of pixels which then outputs a single value for the neuron in the next layer. This operation is realized using a small matrix containing trainable weights. This small matrix is known as the kernel. The aforementioned kernel is then moved over the image in strides, with a stride length of 1 moving the centre of the kernel one pixel across or down until the entire image has been covered. A commonly used kernel size is $3\times3$ pixels, although some networks use larger kernel sizes of up to $11\times11$ pixels early on in the network to reduce input size while the image still contains a high ratio of noise to information ({\sc AlexNet} and {\sc Toothless} make use of this). In later convolutional layers, the kernel is moved across the outputs of the previous layer and not the original image, i.e. the outputs of the previous layer become the input pixels of the current convolutional layer.

This many-to-one mapping during convolution leads to the output of a convolutional layer to be downsampled (i.e. to have a smaller output dimension than the input). 
If downsampling happens too suddenly, this can potentially lead to the loss of too much information without it being incorporated into the model. One workaround for this is padding the input with zeros around the edges, to ensure the output shape is the same size as the input shape. This also assumes a stride length of 1, otherwise downsampling will still occur. This is often referred to as same or zero padding, whereas the absence of padding is known as valid padding. Kernel size, stride length and padding type are all examples of hyper parameters of a CNN, each of which could affect recognition performance.

The output of the final convolutional layer is then flattened into a one dimensional vector, which is then input into the first fully connected layer.

\subsubsection{Max Pooling Layers}
Pooling layers are used to deliberately downsample the input, reducing the input size while preserving the salient features we want the network to learn. Max pooling layers perform downsampling by moving a kernel across the input and returning only the maximum pixel value within a kernel. A max pooling layer's kernel size is normally $2 \times 2$, which halves the input size. 

A reduction in input size is needed so that the the computational requirements of deeper layers  can be reduced or kept constant.

\subsubsection{Training and Backpropagation}

Finding the right combination of weights is referred to as training in Deep Learning terminology, which is done with gradient descent and a loss (or error) function. The loss function represents the error between the target output (the class label that has been provided) and the network's current output (predicted label). A single iteration of training takes place by calculating the gradient of the error function with respect to the weights and biases. The weights and biases are then updated based on the learning rate. However, since only the output layer has a clearly defined target output (the class label), the weights and biases of the intermediate neurons (hidden neurons) cannot be updated in isolation. The amount by which the weights and biases of hidden layers need to be updated depend on all the previous and subsequent layers' parameter values, which makes the gradient descent algorithm computationally expensive (especially on deep networks). This is circumvented by the backward propagation of errors (backpropagation) algorithm \citep{kelley1960}, which calculates the gradient of the final layer's error function, and also reusing partial computations from previous layers, moving ``backwards'' from the output layer through the hidden layers to the input layer.

\subsubsection{Rotational Invariance}
\label{sec:rot_inv}
When a CNN can classify an image irrespective of orientation, it is said to have rotational invariance. CNNs are normally not fully rotationally invariant \citep{Lukic2019b}. Convolutional layers enforce translation equivariance and pooling layers add translation invariance but both of these usually allow only limited invariance to rotations, normally not more than a few degrees \citep{marcos2016}. 
Other Deep Learning architectures are rotationally invariant by design, such as Capsule Networks \citep{sabour2017}. \citet{Lukic2019b} have compared the performance of conventional CNN architectures and capsule networks when they are both used to perform radio galaxy morphological classification.

\subsection{Description of Metrics}
\label{sec:metrics}

\subsubsection{Confusion Matrix and $F_1$-Score}
\begin{figure}
	\includegraphics[scale=0.5]{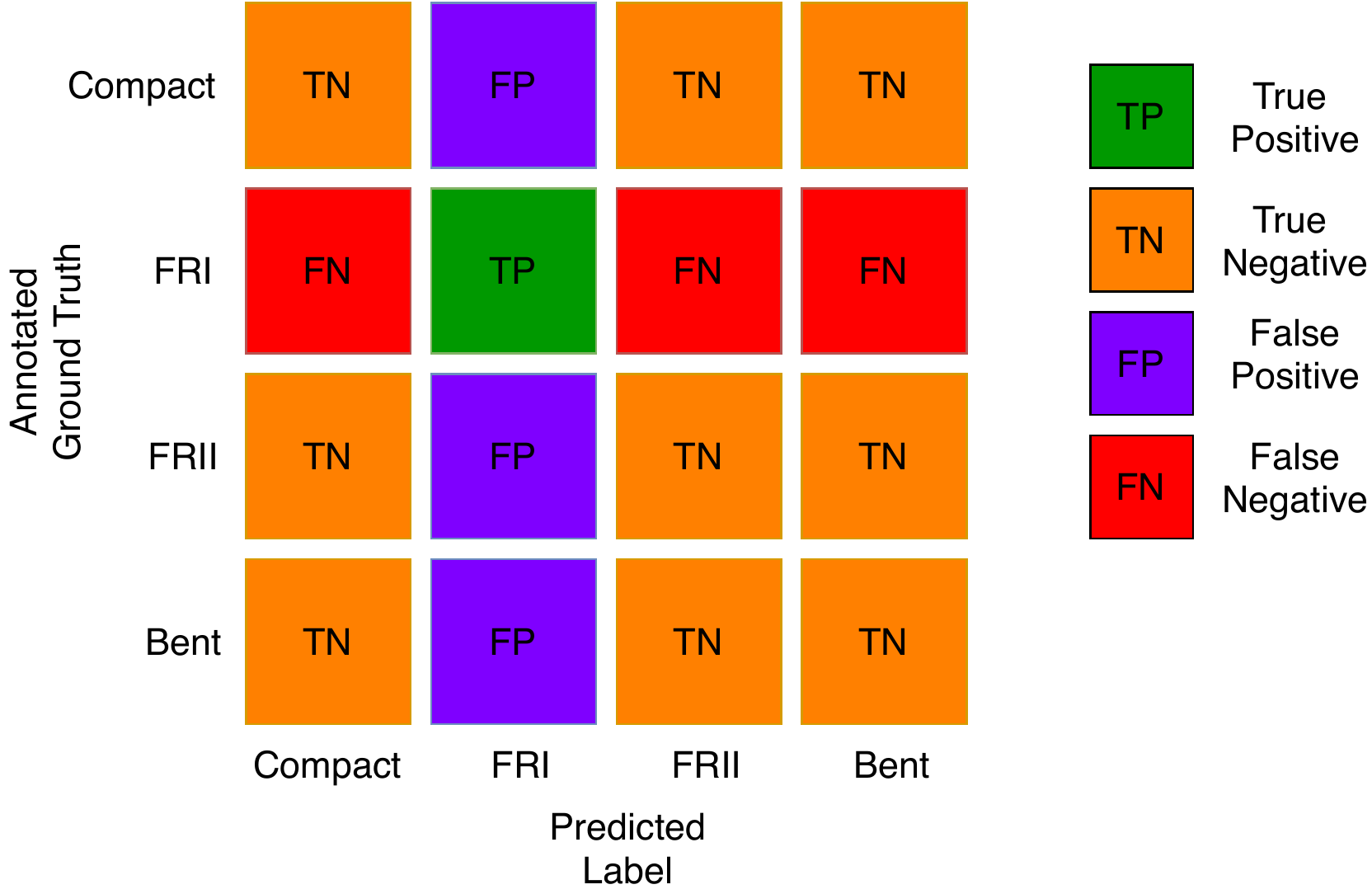}
    \caption{Confusion Matrix Layout. This specific example showcases an assessment of the FRI class.}
    \label{fig:confusion_matrix_layout}
\end{figure}

A useful tool when reporting the results of classifiers is the confusion matrix. The $ij^{\textrm{th}}$ entry of a confusion matrix tells you the number of images that were classified as belonging to class $j$ even though they actually belong to class $i$. In practice, when we depict confusion matrices we often use annotated interpretable labels instead of the aforementioned integer labels. A hypothetical confusion matrix which was obtained after classifying a dataset consisting of radio sources is depicted in Figure~\ref{fig:confusion_matrix_layout}. This confusion matrix depicts how well our classifier could distinguish FRI sources from non-FRI sources. The class for which a classifier's performance is currently being assessed is known as the positive class (the class currently under consideration). The remaining classes are known as the negative classes. In the case of Figure~\ref{fig:confusion_matrix_layout} FRI is our positive class. The depiction in Figure~\ref{fig:confusion_matrix_layout} will of course be different if another class becomes the positive class. Furthermore, the confusion matrix in Figure~\ref{fig:confusion_matrix_layout} also graphically depicts the definition of the following concepts in the case of a multiclass scenario: True Positives, False Positives, True Negatives and False Negatives. The general definition of the above concepts and examples thereof (from Figure~\ref{fig:confusion_matrix_layout}) are listed below: 

\begin{itemize}
    \item True Positives (TP): images belonging to the positive class being classified as such \emph{(sources that were annotated as FRI and classified as such)}.
    \item True Negatives (TN): images from the negative classes that are not classified as belonging to the positive class  (\emph{sources that were annotated as FRII and correctly classified as such or sources that were annotated as Bent, but incorrectly classified as FRII}).
    \item False Negative (FN): images belonging to the positive class not classified as such (\emph{sources annotated as FRI, but incorrectly classified as FRII}).
    \item False Positives (FP): images belonging to the negative classes that are classified as belonging to the positive class (\emph{sources annotated as FRII, but incorrectly classified as FRI}).
\end{itemize}

Recall refers to the ratio of the number of images that were correctly classified as belonging to the positive class to the total number of images in the positive class, i.e.:
\begin{equation}
\textrm{recall} = \frac{\textrm{TP}}{\textrm{TP}+\textrm{FN}}
\end{equation}

Recall is also refered to as the True Positive Rate.
Precision is the ratio of images that were correctly classified as belonging to the positive class to the total number of images that were classified as belonging to the positive class, i.e.:
\begin{equation}
\textrm{precision} = \frac{\textrm{TP}}{\textrm{TP}+\textrm{FP}}
\end{equation}

The weighted average of recall and precision is known as the $F_1$-score:

\begin{equation}
F_1 = 2 \times \frac{\textrm{precision} \times \textrm{recall}}{\textrm{precision} + \textrm{recall}}
\end{equation}

\subsubsection{Mean per Class Accuracy}
Overall accuracy is the ratio of correct classifications for all classses to the total number of samples tested on. With respect to the confusion matrix described in Figure~\ref{fig:confusion_matrix_layout} this would be calculated as the sum of the main diagonal (the TP of each class) divided by the sum of the entire matrix.

Overall accuracy can be a misleading metric, especially when a significant class imbalance is present.

For this purpose we use Mean per Class Accuracy (MPCA), calculated as the mean of the main diagonal of a normalized confusion matrix. The confusion matrix is normalized by dividing each row with the number of samples in that row (which corresponds to the number of samples per class). This metric is less susceptible to class imbalances than overall accuracy. 

\subsubsection{Model Complexity (Trainable Parameters)}

We define model complexity as its number of trainable parameters. The trainable parameters of a neural network are its weights and bias terms. The number of trainable parameters of all model instances associated with a particular architecture remain the same as long as they were created using the same set of hyperparameters.

\subsubsection{Computational Complexity (Floating Point Operations)}
Each architecture's computational complexity is measured using Tensorflow's version 1 profiler \footnote{\url{https://www.tensorflow.org/api_docs/python/tf/compat/v1/profiler/profile}}. The aforementioned profiler measures the number of floating point operations (FLOPs) used by the model in a single forward pass.

\subsubsection{GPU Memory Usage}
Theoretical GPU memory usage was estimated by first determining the memory footprint of the CNNs parameters and then adding to that the amount of active memory the CNN would require when processing a batch of data (a batch size\footnote{Number of samples classified concurrently at any point in time.} of 32 was used in this case). 

\subsubsection{Inference Time and Classification Speed}
Inference time is the time that a CNN requires to classify a single image. Classification speed is the number of images that are classified per second, obtained by inverting inference time. In this paper, inference time was estimated by taking the average of 10 timed runs in which we classified 3072 images (with a batch size of 32).  

\subsubsection{Receptive Field and Effective Stride}
\label{sec:receptive_field}
The final convolutional layer's output is not necessarily the result of a transformation applied to every pixel in the input image (unlike a dense/fully connected layer). Rather, each output pixel has a limited ``field of view'', a limited region in the input image that trickles down through the convolutional layers to become a single output. This is the architecture's theoretical receptive field, as opposed to its  effective receptive field (the pixels in that limited region that had the largest impact on the output) \citep{luo2016}. 

We do not consider the effective receptive field any further in this paper. Moreover, for the sake of simplicity we will refer to the theoretical receptive field simply as the receptive field throughout the remainder of the paper.

Effective stride is defined as the stride between the input layer and the output layer of the convolutional part of a CNN \citep{araujo2019}.\footnote{Effective padding is defined similarly.}

The reader who wants to gain more insight into these topics is referred to \citet{araujo2019}, who provides an in depth description of how the receptive field and the effective stride of a CNN is computed. 

\subsubsection{Data Coverage and Prediction Confidence Threshold}
Data coverage is the percentage of the total dataset that a classifier can assign a label to given a certain confidence threshold. A good characteristic that a classifier should have is that its recognition performance should increase as its prediction confidence threshold is increased. Data coverage will either decrease or remain constant as the prediction confidence threshold is increased, with a significant decrease expected at higher confidence thresholds.
In the ideal case, excluding only a few sources from your dataset will bring about a large gain in recognition performance.

\section{Architectures}
\label{sec:arch}
In this section we discuss the architectures we considered for our study. We also present auxiliary useful information that will improve the reader's understanding of the paper.

\subsection{Models vs Architectures}
The terms architecture and model are not used interchangeably in the context of this study. We refer to an architecture as the layout of the network's structure, whereas a model is a trained instance of the architecture. Models of the same architecture are differentiated by the data it was trained on and other hyperparameters such as different learning rates or the optimizer used during training. 

\subsection{List of Architectures}

\begin{figure*}
	\includegraphics[width=0.47\textwidth]{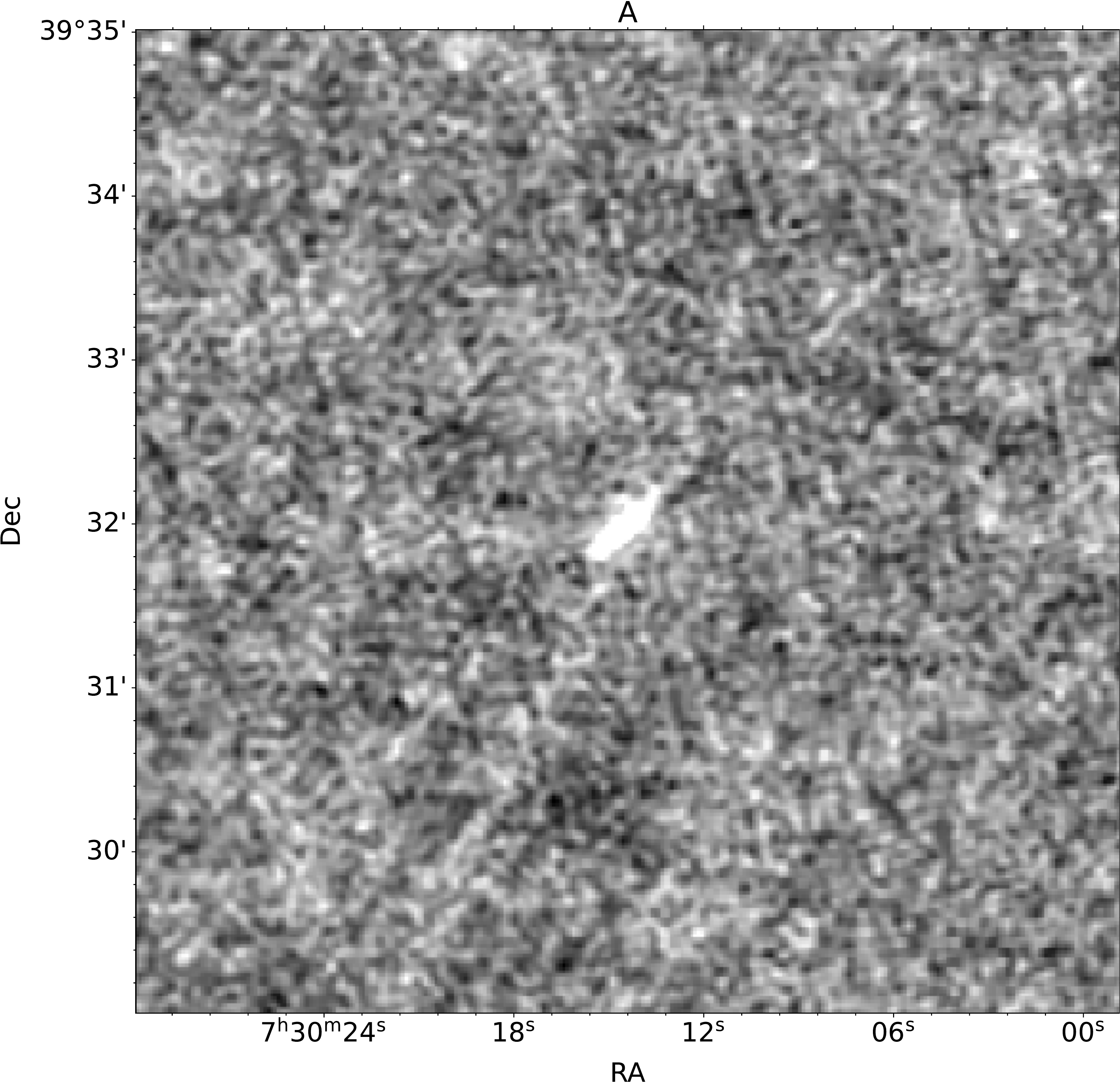}
	\includegraphics[width=0.47\textwidth]{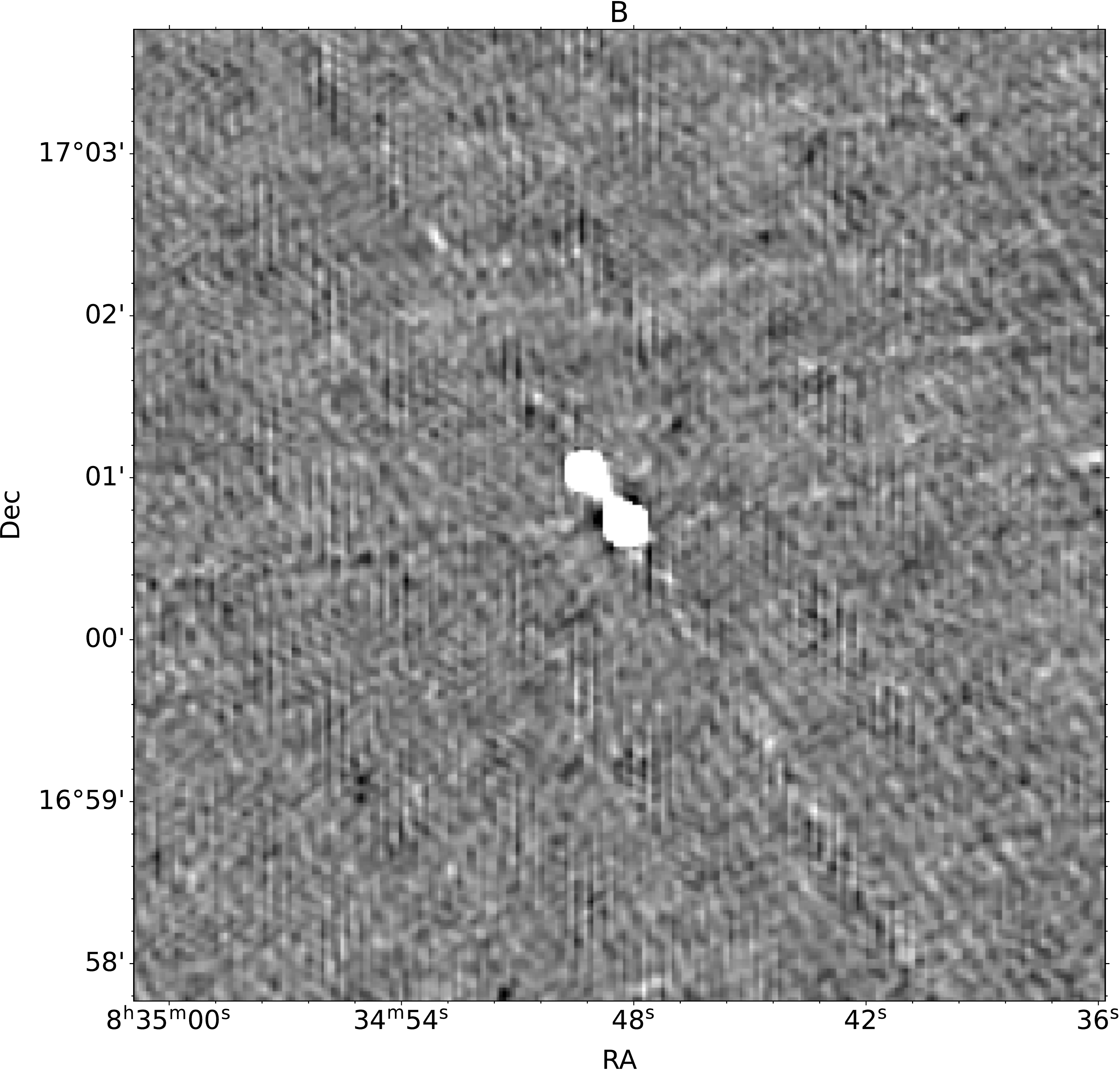}
	\includegraphics[width=0.47\textwidth]{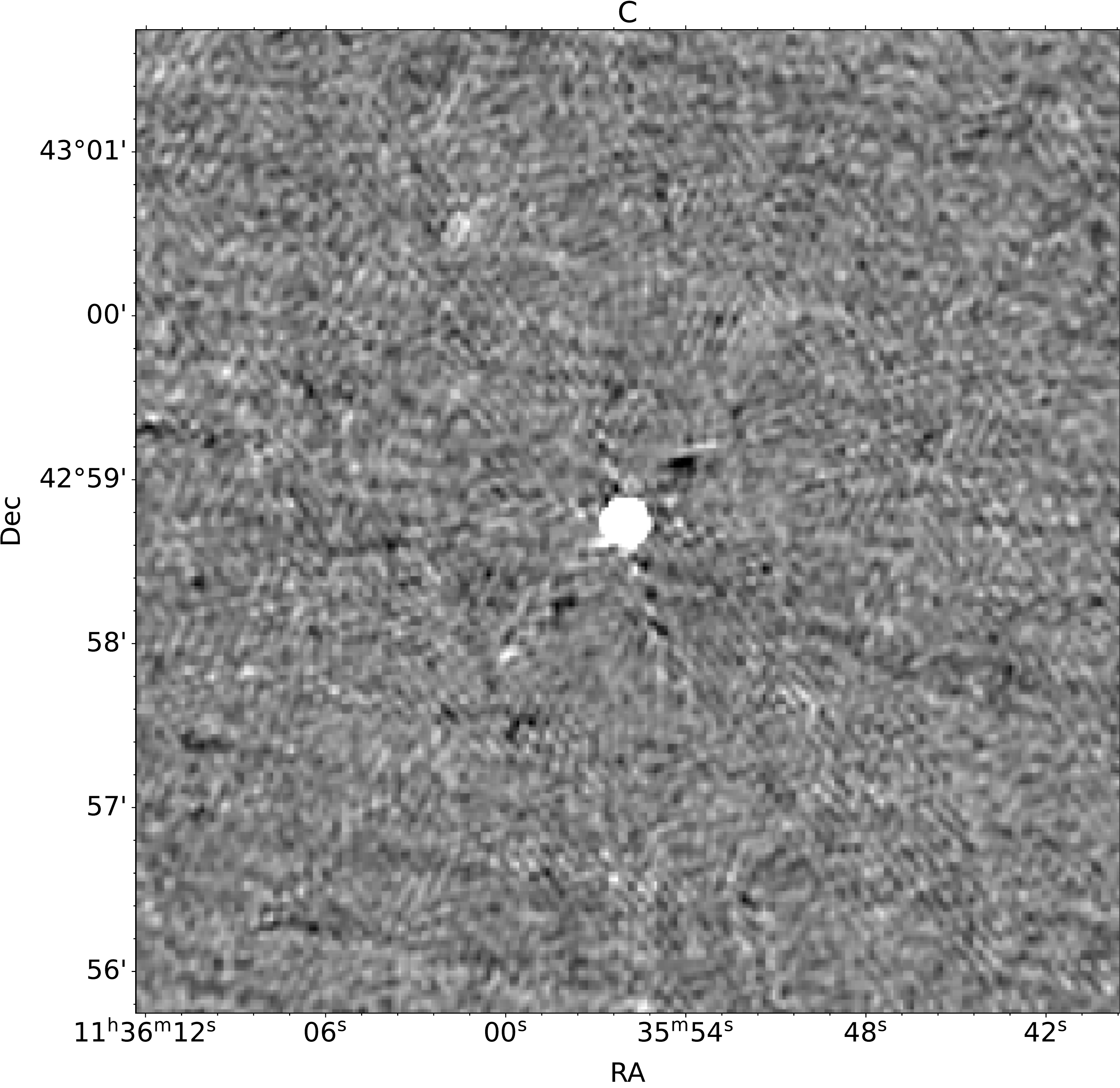}
	\includegraphics[width=0.47\textwidth]{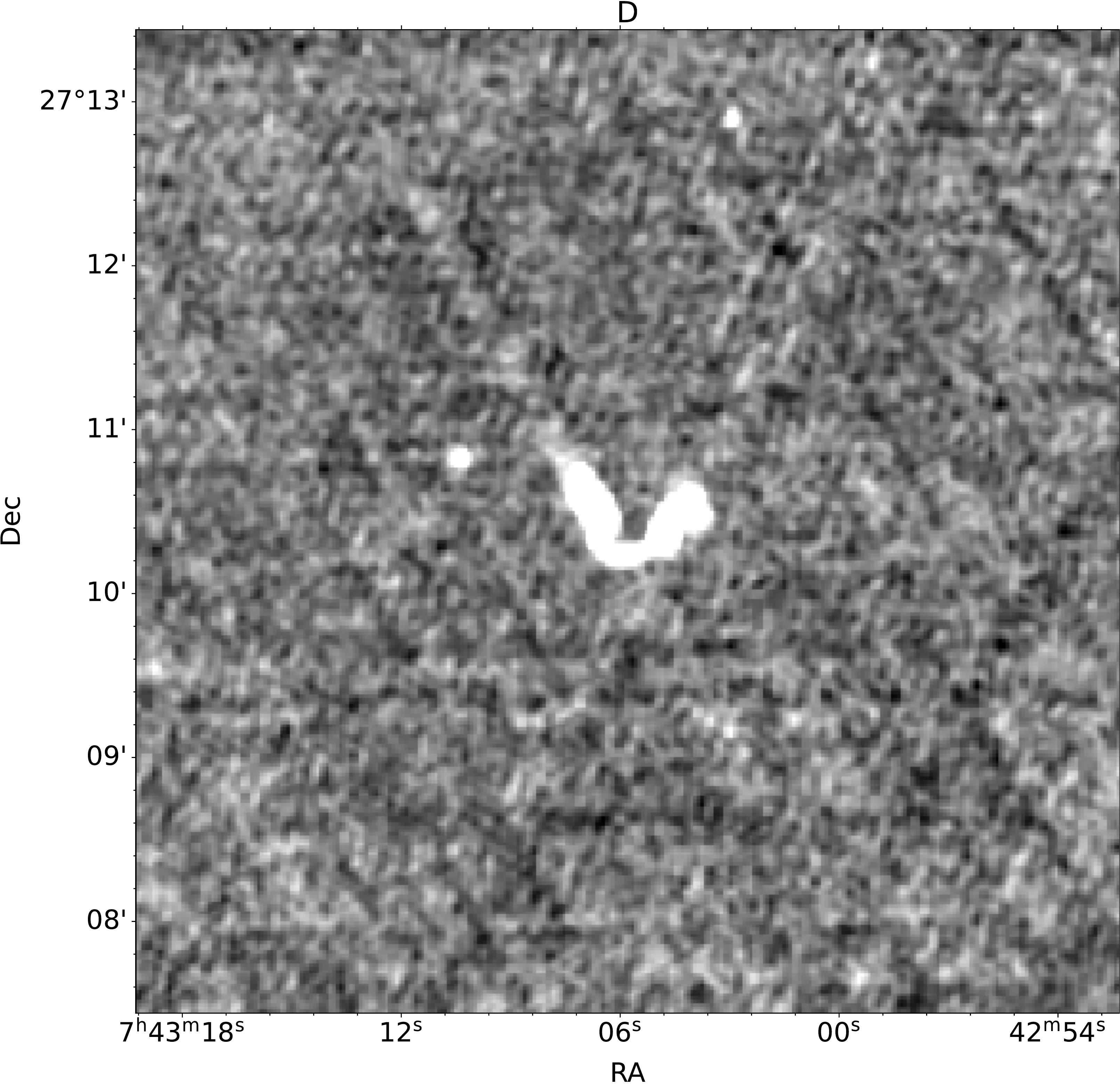}
    \caption{Examples of the different radio morphologies that have been classified in this study. Examples include an FRI (A), an FRII (B), a compact radio source (C) and a bent-tailed radio galaxy (D). All examples are shown before the preprocessing step.}
    \label{fig:Examples}
\end{figure*}

\begin{table}
\caption{List of architectures and their keys for all figures. Architectures marked $\dagger$ have been modified from their original form.}
\label{tab:arch_keys}
\begin{tabular}{lll}
\hline
Architecture Name          & Key   & Study \\
\hline
{\sc AlexNet}             & ALN     & \citet{Krizhevsky2012}\\
{\sc ATLAS X-ID}$\dagger$   & ATL     & \citet{Alger2018}\\
{\sc ConvNet4}            & CN4    & \citet{Lukic2019a}\\
{\sc ConvNet8}            & CN8    & \citet{Lukic2019a}\\
{\sc FIRST Classifier}    & 1stC     & \citet{Alhassan2018}\\
{\sc FR-Deep}             & FR-D     & \citet{Tang2019}\\
{\sc Hosenie}             & H     & \citet{Hosenie2018}\\
{\sc MCRGNet}$\dagger$            & MCRG     & \citet{Ma2019}\\
{\sc Radio Galaxy Zoo}    & RGZ     & \citet{Lukic2018}\\
{\sc SimpleNet}           & CNs    & \citet{Lukic2019a}\\
{\sc Toothless}$\dagger$           & TLS     & \citet{Aniyan2017}\\
{\sc CLARAN}$\dagger$ ({\sc VGG16D})    & VGG     & \citet{Wu2019}\\
{\sc ConvXpress}          & CXP       &  Novel\\
\hline
\end{tabular}
\end{table}
The architectures we considered in our comparison are listed in Table~\ref{tab:arch_keys}, providing the architecture names, the corresponding studies from the literature as well as shortened keys assigned for use in plots (see Figure~\ref{fig:ball_plot} as an example). Some studies have contributed more than one architecture \citep{Lukic2019a}. All of the architectures listed in Table~\ref{tab:arch_keys} were modified to enable them to discern between four types of radio sources (i.e. the number of output classes were changed to four). Example images of the four classes that we consider in this paper are depicted in Figure~\ref{fig:Examples}. 

\subsubsection{Unmodified Architectures}

{\sc AlexNet}, {\sc ConvNet4}, {\sc ConvNet8}, {\sc FIRST Classifier}, {\sc FR-Deep}, {\sc Hosenie}, {\sc Radio Galaxy Zoo} and {\sc SimpleNet} were not modified in any further way.

\subsubsection{Modified Architectures}
\label{sec:modifications}
The following architectures were further modified:
\begin{itemize}
   \item {\sc ATLAS X-ID}: {\sc ATLAS X-ID} was not designed explicitly for radio galaxy classification, but rather for finding host galaxies for radio sources by cross identification. The CNN described in the paper had an additional input vector of 10 features from the candidate host in the SWIRE survey, which has not been included in the modified version.
   \item {\sc MCRGNet}: has been adapted from the neural network described by \citet{Ma2019}. Initially this network was pretrained as the encoder level of a Convolutional Auto-Encoder using an unlabelled sample and then fine-tuned on a labelled sample. Several of these CNNs would be combined to form a dichotomous tree classifier, each classifying a subset of the classes. Due to computational constraints, the pretraining step has been left out and only a single instance of this architecture is used.
   \item {\sc Toothless}: originally implemented as a fusion classifier consisting of 3 binary classifiers, classifying either FRI/FRII,FRI/Bent and FRII/Bent respectively. If two classifiers would predict a source as the same class with a 60\% probability, the classification would be accepted, if both predicted with less than the 60\% confidence threshold, a `?' would be appended to the classification. Additionally, should none of the classifiers give the same class output, the source is labelled as "Strange". To reduce computational requirements, only a single classifier instance is considered.
   \item {\sc CLARAN}: {\sc CLARAN} takes as input a radio source and a corresponding infrared image, after which it outputs a bounding box showing the location and size of the detected radio source. The source morphology is given in the format $iC\_jP$, where $i$ is the number of components and $j$ is the number of flux-density peaks. A corresponding probability of the morphology is also output. {\sc CLARAN} uses {\sc VGG16D} \citep{Simonyan2014} as a classification layer that is fed into a region of interest classifier. While the entire architecture was not suitable for this study, the {\sc VGG16D} classifier layer was appropriate to include. Note that the {\sc VGG16D} architecture we include in this study, in contrast with {\sc CLARAN}, can only assign one label to each image it receives and would, therefore, not fare well if the images it receives contain multiple source.
\end{itemize}

\subsection{Impact of Modifications}

At this point in time we should take a moment to consider the potential impact that the modifications we discuss in the beginning of Section~\ref{sec:arch} and those in Section~\ref{sec:modifications} will have on the recognition performance of the architectures presented in the studies from Table~\ref{tab:arch_keys}. But first, it should be duly noted that the proposed modifications are a necessity as these modifications make it possible to perform a meaningful comparison of these architectures. Three major modifications were discussed in the beginning of Section~\ref{sec:arch} and in Section~\ref{sec:modifications}:

\begin{description}

\item[\textbf{Output Classes}] The number of the output classes and in some cases even the output-labels of the output classes were altered ({\sc Toothless} as an example of the former, {\sc CLARAN} as an example of the latter). This alteration, however, is standard practise within the field of Deep Learning. Take {\sc AlexNet} as an example it was originally designed for the ImageNet Challenge, but it is nowadays used to solve many other types of image recognition problems (i.e. the number of classes and the output-lables it can produce differ from its original use case). Generally speaking, if a CNN architecture is identified that can discern between $N$ different classes, then its recognition performance will normally not deteriorate significantly if the number of classes that one considers is either reduced or increased by one (given that it is properly re-trained). Moreover, neither would considering $N$ completely different labels have a significant impact on its performance. There are of course exceptions to this, if the nature of the problem is changed completely or the inherent separability of the dataset changes significantly this generalization might not necessarily remain true. 
\item[\textbf{Architecture Instances}] Only single architecture instances were considered (as an example only a single architecture instance of {\sc Toothless} was used). Multiple instances of any architecture can be incorporated into a more complex classifier (like a fusion classifier). This will certainly improve the recognition performance of a particular architecture. However, knowing how a single instance of the architecture performs enables us to identify which architectures will ultimately perform better if they are chosen to create a more complex classifier.
\item[\textbf{Data}] The same dataset (no peripheral data was included in our experiments so that a fair comparison between the architectures in Table~\ref{tab:arch_keys} could be made even though additional data would have resulted in improved performance of certain architectures) was used to evaluate each architecture. As mentioned in Section~\ref{sec:modifications} {\sc MCRGNet} and {\sc ATLAS X-ID} is particularly affected by this.  
\end{description}

\subsubsection{Novel Architecture}
The architecture of {\sc ConvXpress} is based on the architecture of {\sc ConvNet8} and {\sc VGG16D}. {\sc ConvXpress} is deeper than {\sc ConvNet8} (11 vs 8 convolutional layers) and uses the convolutional stack structure introduced by {\sc VGG16D}. Each stack is comprised of 3 convolutional layers (except for the last stack, which is only two) and a max pooling layer. This was developed to match or enlarge the receptive field size (see Section~\ref{sec:receptive_field}) of {\sc AlexNet}'s convolutional layers without having to use {\sc AlexNet}'s large kernel size. The effective receptive field of {\sc ConvXpress}'s first convolutional stack is $11 \times 11$, the same size as {\sc AlexNet}'s first convolutional layer. However, the stack structure has applied three non-linear rectification functions compared to {\sc AlexNet}'s single activation, making the model more discriminative \citep{Simonyan2014}. 

In addition to this, the number of parameters required are reduced by stacking: a layer with an $11 \times 11$ kernel with $C$ input channels require $ 11^2C^2=121C^2 $ parameters, while 3 layers with a $3 \times 3$ kernel and $C$ input channels require only $3 (3^2C^2) = 27C^2 $ parameters. 

{\sc ConvXpress} has a non-standard stride length, similar to {\sc MCRGNet}, {\sc Toothless}, {\sc AlexNet} and {\sc Radio Galaxy Zoo}. In particular, it makes use of a stride length of 2 in the first convolutional layer of the first and second convolutional stacks. 
The Dense (or fully-connected) layers are the same as {\sc ConvNet8}'s, using a linear activation in the second last layer with an L2 kernel regularizer. {\sc ConvXpress} also contains five dropout layers. During each training step a dropout layer randomly turns some of the neurons in the layer that comes before it off (i.e. it blocks their output from propagating to the next layer). The probability that a specific neuron is turned off is known as the dropout rate $p$. This is similar to creating and training many small networks within the larger network \citep{Srivastava2014}. The value for $p$ for all the dropout layers in {\sc ConvXpress} is 0.25.  The only exception is the last dropout layer. For the last layer the value of $p$ is equal to 0.5.

The architecture of {\sc ConvXpress} is presented in Table~\ref{tab:CXP}.

\begin{table}
\caption{{\sc ConvXpress} Architecture Layout}
\label{tab:CXP}
\begin{tabular}{lrrrl}
\hline
Layer        & Depth & Kernel Size & Stride Length & Activation \\ \hline
Conv2D       & 32    & 3           & 2             & ReLU       \\
Conv2D       & 32    & 3           & 1             & ReLU       \\
Conv2D       & 32    & 3           & 1             & ReLU       \\
MaxPooling2D &       & 2           & 1             &            \\
Dropout      &       &             &               &            \\
Conv2D       & 64    & 3           & 2             & ReLU       \\
Conv2D       & 64    & 3           & 1             & ReLU       \\
Conv2D       & 64    & 3           & 1             & ReLU       \\
MaxPooling2D &       & 2           & 1             &            \\
Dropout      &       &             &               &            \\
Conv2D       & 128   & 3           & 1             & ReLU       \\
Conv2D       & 128   & 3           & 1             & ReLU       \\
Conv2D       & 128   & 3           & 1             & ReLU       \\
MaxPooling2D &       & 2           & 1             &            \\
Dropout      &       &             &               &            \\
Conv2D       & 256   & 3           & 1             & ReLU       \\
Conv2D       & 256   & 3           & 1             & ReLU       \\
MaxPooling2D &       & 2           & 1             &            \\
Dropout      &       &             &               &            \\
Flatten      &       &             &               &            \\
Dense        & 500   &             &               & Linear     \\
Dropout      &       &             &               &            \\
Dense        & 4     &             &               & Softmax   \\
\hline
\end{tabular}
\end{table}

\subsubsection{Excluded Architectures}
Some architectures were excluded from the study for either being originally designed to perform a task other than classification or in order to restrict the scope of the study to conventional CNN architectures. The following architectures were excluded:
\begin{itemize}
      \item Convosource: Designed for source-finding, to extract the pixels that belong to an astronomical source from an image with background noise \citep{Lukic2019b}.
  \item COSMODEEP: Designed to perform a combination of source finding and classification by breaking up larger images into smaller tiles that are then individually classified as either containing no signal or containing a radio source \citep{Gheller2018}.
  \item DEEPSource: Aimed at source-finding in low signal-to-noise ratio cases \citep{Sadr2019}.
  \item Capsule Networks: This study limits the focus of comparison to conventional CNN architectures described in the literature. \citet{Lukic2019b} compared Capsule Network performance with conventional CNN architectures.
\end{itemize}

\subsubsection{{\sc AlexNet} and {\sc Toothless}}

Just as {\sc AlexNet} has been a seminal work in image classification for CNNs, so too has {\sc Toothless} \citep{Aniyan2017} made its mark on the classification of radio galaxies for being the first CNN developed for this very purpose. It has thus been referenced in almost all of the subsequent works listed in Table~\ref{tab:arch_keys}. {\sc Toothless} is based on {\sc AlexNet}'s architecture and does not differ much other than the type of padding used, with the original {\sc AlexNet} design using valid padding (no padding is applied around the edges of the input of a layer) rather than same padding (zero padding around the edges of the input of a layer to ensure there is no size reduction other than that caused by stride length). This small difference does have a slight effect on performance, since the size of the input is being reduced steadily with valid padding, less computational resources are required for {\sc AlexNet} than for {\sc Toothless}. The type of padding at different layers is something that should be carefully considered when designing an architecture, since this might shrink input too fast, throwing away useful information. Same or zero padding will work better for a wider range of input resolutions.

\section{Data Description}
\label{sec:data}

\citet{Ma2019} used two datasets in their study: 
\begin{itemize}
    \item \textbf{the Labelled Radio Ralaxy (LRG) dataset:} a curated dataset that contain well attested sources from the CONFIG \citep{Gendre2008,Gendre2010}, GROUPS \citep{Proctor2011}, FRICAT \citep{Capetti2017a}, FRIICAT \citep{Capetti2017b}, FR0CAT \citep{Baldi2018} and \citet{Cheung2007} catalogues (see Table~\ref{tab:catalogues}). The dataset contains 1442 sources and consists of 6 classes (Compact, FRI, FRII, Bent, X-shaped and Ringlike)\footnote{The total number of sources reported in \citet{Ma2019} and the total number of sources in the catalog available on their Github repo differ slightly.}. 
    \item \textbf{the Unlabelled Radio Galaxy (URG) dataset:} a dataset containing 14245 AGNs from the Best-Heckman sample \citep{Best2012}. This dataset was manually labelled by \citet{Ma2019}. 
\end{itemize}

We use slightly modified versions of these two datasets in our study. We will refer to the modified version of the LRG dataset as the Modified Labelled Radio Ralaxy (MLRG) dataset throughout the rest of the paper (it contains 1,328 sources). Similarly we will refer to the modified URG dataset as the Modified Unlabelled Radio Galaxy (MURG) dataset (it contains 14,093 sources). We made the following modifications. We removed all X-shaped and Ringlike sources from both the LRG and the URG. We also removed error-prone images from the URG dataset (i.e. images consisting only of \texttt{NaN} values)\footnote{Three Compact, two FRI and two FRII sources were also removed.}. Moreover, all FR0 sources were added to the Compact class. 
\begin{table}
\caption{Class breakdown per catalogue for the LRG dataset}
\setlength{\tabcolsep}{3pt}
\label{tab:catalogues}
\centering
\begin{tabular}{lrrrrrrr}
\hline
               & Compact & FR0 & FRI & FRII & Bent & X  & Ring \\ \hline
CoNFIG         & 270     & 8   & 14  & 350  & 9    & 0  & 0    \\
FR0CAT         & 0       & 104 & 0   & 0    & 0    & 0  & 0    \\
FRICAT         & 1       & 19  & 173 & 0    & 5    & 0  & 0    \\
FRIICAT        & 0       & 0   & 0   & 80   & 8    & 3  & 0    \\
Proctor (2011) & 0       & 1   & 0   & 0    & 284  & 0  & 32   \\
Cheung (2007)  & 0       & 2   & 0   & 0    & 0    & 79 & 0    \\ \hline
Total          & 271     & 134 & 187 & 430  & 306  & 82 & 32    \\ \hline
\end{tabular}
\end{table}

The final class breakdown of both datasets are presented in Table~\ref{tab:data_description}. 

The final catalogues we used are partially presented in Tables~\ref{tab:mMLRG} and \ref{tab:mMURG}. The rest of these catalogues are available on our Github repository. A script that can download the sources from the catalogues is also provided on our Github repository. It downloads FIRST cutouts \citep{becker1995} in FITS format (300 by 300 pixels) via the Skyview tool \citep{mcglynn1998}.

\begin{table}
\setlength{\tabcolsep}{2.3pt}
\caption{The first 5 rows of the MLRG sample, the full table is available on the projects \href{https://github.com/BurgerBecker/rg-benchmarker}{Github Repository}.}
\label{tab:mMLRG}
\centering
\begin{tabular}{lrrr}
\hline
Source & Right Ascension & Declination & Classification  \\
Name & (degrees) & (degrees) &  \\
\hline
J000330.73+002756.1 & 0.05854 & 0.46558         & Bent-tailed              \\
J001247.57+004715.8 & 0.21321 & 0.78772         & FRII             \\
J002107.62-005531.4 & 0.35212 & -0.92539        & FRII            \\
J002900.98-011341.7 & 0.48361 & -1.22825       & Compact             \\
J003930.52-103218.6 & 0.65848 & -10.5385 & FRI        \\
\hline
\end{tabular}
\end{table}

\begin{table}
\setlength{\tabcolsep}{2.3pt}
\caption{The first 5 rows of the MURG sample, the full table is available on the projects \href{https://github.com/BurgerBecker/rg-benchmarker}{Github Repository}.}
\label{tab:mMURG}
\centering
\begin{tabular}{lrrrrr}
\hline
Source & Right Ascension & Declination & Classification  \\
Name & (degrees) & (degrees) &  \\
\hline
J000001.57-092940.3 & 0.00044 & -9.49453 & Compact \\
J000025.55-095752.8 & 0.00710 & -9.96467 & FRI \\
J000027.89-010235.4 & 0.00775 & -1.04317 & Compact \\
J000049.32-005042.9 & 0.01370 & -0.84525 & FRI \\
J000052.92+003044.6 & 0.01470 & 0.51239  & FRII \\
\hline
\end{tabular}
\end{table}

\begin{table}
\caption{Comparison of the MLRG and MURG datasets we use in this study.}
\label{tab:data_description}
\centering
\begin{tabular}{lll}
\hline
Class   & MLRG  & MURG \\ \hline
Compact & 405  & 6093  \\
FRI     & 187  & 5039  \\
FRII    & 430  & 2072  \\
Bent    & 306  & 889   \\ \hline
Total   & 1328 & 14 093 \\ \hline
\end{tabular}
\end{table}

\section{Experimental Setup}
\label{sec:experiments}
This section describes the experimental setup we used. The image prepossessing procedure we adopted is described in Section~\ref{sec:imgProc}. The hardware used as well as other important overarching experimental information is presented in Section~\ref{sec:prel}. The two main experiments conducted are described in Section~\ref{sec:overfit} and Section~\ref{sec:MURGexperiment}. We end this section by describing the ensemble classifiers that we constructed.   

\subsection{Image Preprocessing}
\label{sec:imgProc}
The preprocessing steps used are: the images were first normalized and then thresholding was applied. Allowing some noise in the training data improves learning for very deep networks \citep{neelakantan2015}. The thresholding method used assigns a zero value to all pixels with a value below the threshold of three standard deviations above the mean pixel value of the specific image. Otherwise, the pixel value is kept the same.

\subsection{Preliminaries}
\label{sec:prel}
All training was performed on a Nvidia Tesla V100 32GB. Our architectures were constructed using the Deep Learning framework: Keras \citep{chollet2015}. To ensure replicable results we provided a random seed to all non-deterministic processes. In addition to this step, Tensorflow requires you to set the \texttt{TF\_CUDNN\_DETERMINISTIC} environment variable to `1' or `true' (alternatively, depending on the version of Tensorflow being used, either the Nvidia Tensorflow-Determinism patch can be applied or the \texttt{TF\_DETERMINISTIC\_OPS} environment variable must be set \footnote{\url{https://pypi.org/project/tensorflow-determinism/}}). A customized version of the Keras Data Generator class was used to load images during training, validation and testing.

Each architecture was trained for 16 epochs with a learning rate dependant on the number of parameters in the architecture. Adam was used as the optimizer with a callback function that reduces the learning rate once a loss plateau is reached. This callback function reduces the learning rate by half after 3 epochs in which the validation loss has not decreased by the set threshold: 0.001 (i.e. this is the learning rate scheduler we used). At the end of each epoch, another callback function assesses the validation loss. If the current model has a lower validation loss than the previous lowest validation loss, this model is saved and the previous one is discarded. This model is used to represent the architecture in the tests of both experiments below. This process is necessary to prevent overfitting by storing the models that generalize well on the validation data.

Furthermore, the experiments below were repeated three times  (different seed values for the random processes and the subset selection were assigned during each of the three runs). Using different seed values results in a different weight initialization for the CNNs and a different subset selection for the training, validation and testing sets during each run. This is done to get a more accurate representation of the architecture's performance for the chosen hyperparameters and to assess the validity of the results.

\subsection{Overfit Experiment}
\label{sec:overfit}

In this experiment we emulate the type of training most of the architectures in Table~\ref{tab:arch_keys} employed in their respective studies: training, validation and testing on a small curated dataset. More specifically: the architectures are trained on a subset of the MLRG dataset and then tested on a mutually exclusive subset of the MLRG dataset. The architectures are then re-tested on the full MURG dataset. The training, validation and test set breakdown used for this experiment are summarized in Table~\ref{tab:MLRG_test}. We elaborate further in this regard in the sections that follow.

\subsubsection{Augmentation}
The training and validation sets are augmented by rotating each source at 15 degree intervals, leading to 24 rotated samples of each image. The total number of augmented samples are given in parentheses in Table~\ref{tab:MLRG_test}. This is done to increase the number of samples for validation and training as well as addressing rotational invariance (discussed in section~\ref{sec:rot_inv}).

Each source is rotated after preprocessing and then saved as a new FITS image with the rotation factor added to the original file name.

\subsubsection{Training and Validation}
The training and validation data are selected from the MLRG dataset (the samples in these datasets depend of the chosen seed value and as such differ for each experimental run). Training is performed on 80 unique sources per class (1920 after augmentation) and validation on 60 unique sources per class (1920 after augmentation). Using the aforementioned number of sources for training and validation allows for testing on roughly 25\% of the smallest class (FRI) in the MLRG dataset. The split ratio's are in line with both a standard training/validation/test split commonly used in practice and with the splits used in most of the studies of the architectures presented Table~\ref{tab:arch_keys}.

\subsubsection{Testing}
We first test the resulting models on the test split of the MLRG dataset. The models are then tested on the full MURG dataset. Both sets are given in Table~\ref{tab:MLRG_test}.

\subsection{MURG Random Split Experiment}
\label{sec:MURGexperiment}
In this section we describe an experiment that is designed to be less susceptible to overfitting. While the MLRG sample provides excellent examples of each class, the application of stringent selection criteria results in the loss of samples that could make a model more robust. Similar to allowing some noise to remain after preprocessing an image to facilitate better training, allowing samples in the training set that are less-than-perfect examples provides a more nuanced understanding of the class distinction and can lead to more robust classification systems. This problem might be more adequately addressed by taking into account the annotator's confidence in their classification, as a separate input in the dense layer for example, however we leave this for future exploration. 

While the aforementioned issue is worth noting, the small size of the MLRG sample is the most serious concern when it comes to potentially overfitting a model. Large training sets are required to train any type of Deep Neural Network. The MURG sample provides a dataset that is large enough to be used to train a Deep Learning model. Although the training sample used for the MURG random split experiment is relatively small compared to what other deep learning studies have used, there is a 312.5\% increase in the size of the training set used for this experiment when its size is compared with the size of the training set used during the Overfit experiment. The results of the MURG random split experiment provides a more realistic reflection of the architectures in Table~\ref{tab:arch_keys}'s expected performance when deployed in practice. 

More specifically: for this experiment architectures are trained on a subset of the MURG sample and tested on a test split of the MURG sample. The training, validation and test set breakdown used for this experiment are summarized in Table~\ref{tab:MURG}. We elaborate further in this regard in the sections that follow. 

\subsubsection{Augmentation}

The training set is augmented by rotating each source at 15 degree intervals, leading to 24 rotated samples of each image. This results in 24,000 and 9,600 sources for training and validation respectively after augmentation, as shown in parentheses in Table~\ref{tab:MURG}.

\subsubsection{Training and Validation}
The training and validation data are sampled from the MURG dataset. Training is performed on a random selection of 250 sources per class (6000 after augmentation) and validation on 100 sources per class (2400 after augmentation). Again, exactly which sources are selected is determined by the random seed that was used during each experimental run. While this is a much larger training and validation split than what is used for the Overfit experiment, it is significantly smaller than what is normally used when training a CNN, the total training set makes up only 7.87\% of the total dataset compared to normally selecting closer to 50\% of the set or more for training. This smaller selection was chosen to assess the efficacy of model generalization when training on a relatively small subset of the data. 

\subsubsection{Testing}

Testing was performed on a MURG test split. The total number of samples in this test set is given in Table~\ref{tab:MURG}.

\begin{table}
\caption{Overfit Experiment: Training, validation and test set break down per class. Note the two different sets which are used for testing which relate to the results in Figure~\ref{fig:overfit}. The values in parentheses are the number of augmented samples.}
\label{tab:MLRG_test}
\centering
\begin{tabular}{lrrrr}
Class   & Training Set      &   Validation Set  & MLRG      &  MURG\\
        & (Augmented)       &   (Augmented)  & Test Set      &  Test Set\\ \hline
Compact &   80  (1920)         &   60  (1440)         &   265             &   6093\\
FRI     &   80  (1920)         &   60  (1440)         &  47               &   5039\\
FRII    &   80  (1920)         &   60  (1440)         &  290              &   2072\\
BENT    &   80  (1920)         &   60  (1440)         &  166              &   889\\ \hline
Total   &   320  (7680)       &   240  (5760)       &  768              &   14 093\\ \hline
\end{tabular}
\end{table}

\begin{table}
\caption{MURG Random Split Experiment: Training, validation and test set break down per class. The test set is a subset of the MURG samples. The values in parentheses are the number of augmented samples.}
\label{tab:MURG}
\centering
\begin{tabular}{lllr}
Class   & Training Set  &   Validation Set                  &   Test Set\\
        & (Augmented)  &   (Augmented)                      &           \\ \hline
Compact &   250 (6000)       &   100 (2400)                 &   5743\\
FRI     &   250 (6000)        &   100 (2400)                &   4689\\
FRII    &   250 (6000)        &   100 (2400)                &   1722\\
BENT    &   250 (6000)        &   100 (2400)                &   539\\ \hline
Total   &   1000 (24 000)       &   400 (9600)           &   12 693\\ \hline
\end{tabular}
\end{table}

\subsection{Ensemble Method Classifier}
\label{sec:ensemble}
An ensemble classifier is a combination of several classifiers trained to perform the same purpose (such as classification). These ensembles often generalize better than any of their constituents and are less likely to have the same pitfalls as their constituents (overfitting to a specific class for example). Two ensemble classifiers have been created from the MURG Random Split Experiment: 

\begin{enumerate}
    \item Ensemble of all classifiers (ENA)
    \item Top 4 Classifier Ensemble, selected based on their MPCA. From here on referred to as the SKA Artificial Intelligence Network (SKAAI Net in Figure~\ref{fig:ensemble} or as SKN in Figure~\ref{fig:f1_score}).
\end{enumerate}

Both ensembles sum the output probabilities of their constituent classifiers and take the highest probability as the output class.

\section{Results: Overfit Experiments}
\label{sec:overfit_results}
Section~\ref{sec:overfit_acc} reports only on the results form the Overfit Experiment (see Section~\ref{sec:overfit}). In Section~\ref{sec:overfit_vs_MURG}, we compare the results obtained from the Overfit experiment with that of the MURG Random Split Experiment (see Section~\ref{sec:MURGexperiment}). 

\subsection{Overfit Accuracy}
\label{sec:overfit_acc}
\begin{figure}
	\includegraphics[scale=0.35]{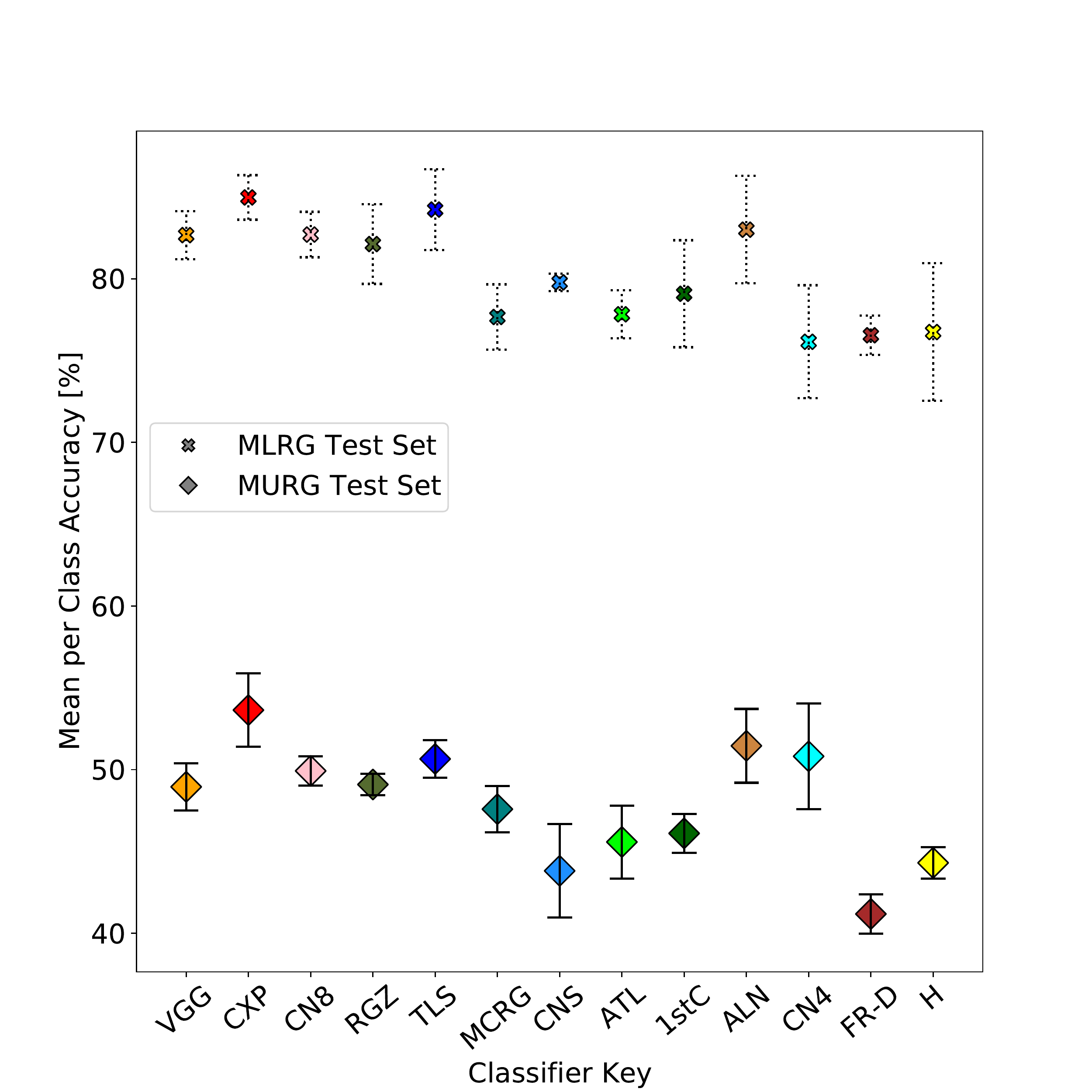}
    \caption{Models trained on the MLRG sample in the ``Overfit'' experiment, reporting MPCA averaged over three iterations with different random seed values. The crosses represent each architectures's averaged MPCA on the MLRG sample's test split, while the diamonds show averaged MPCA on the full MURG dataset. The results show that training and testing architectures on a small sample, such as the MLRG sample, can give misleading expectations for performance on a larger dataset that has not been curated as thoroughly, such as the MURG dataset. The standard deviation for each architecture is also shown.}
    \label{fig:overfit}
\end{figure}

Figure~\ref{fig:overfit} shows the averaged MPCA over three runs of the Overfit Experiment for each architecture. The data-points depicted by the cross markers represent the MPCA results associated with the MLRG test set (described in Table~\ref{tab:MLRG_test}), while the data-points depicted by the diamond markers represent the MPCA results associated with testing on the full MURG dataset. All models experience a more than 20\% decrease in performance when switching from the MLRG test set to the MURG dataset, clear evidence of the models overfitting to the MLRG training set and the results being unreliable to assess performance on a larger dataset. Given that the training sample size is only 2.2\% that of the test sample size, model performance on the MURG data set is not as bad as one would expect. The results, however, indicate that the models trained on only the MLRG set should not be used for autonomous classification in practice, since for the MURG dataset most of the models have a sub 50\% accuracy in at least one class.

Training and testing models on small datasets  give a skewed perception of architecture performance. Models need to be trained and assessed on samples that are representative enough of the underlying data distribution that underpins the classification problem at hand (a too small training dataset prevents this). 

\subsection{Overfit vs MURG Random Split Results}
\label{sec:overfit_vs_MURG}
\begin{figure}
	\includegraphics[scale=0.35]{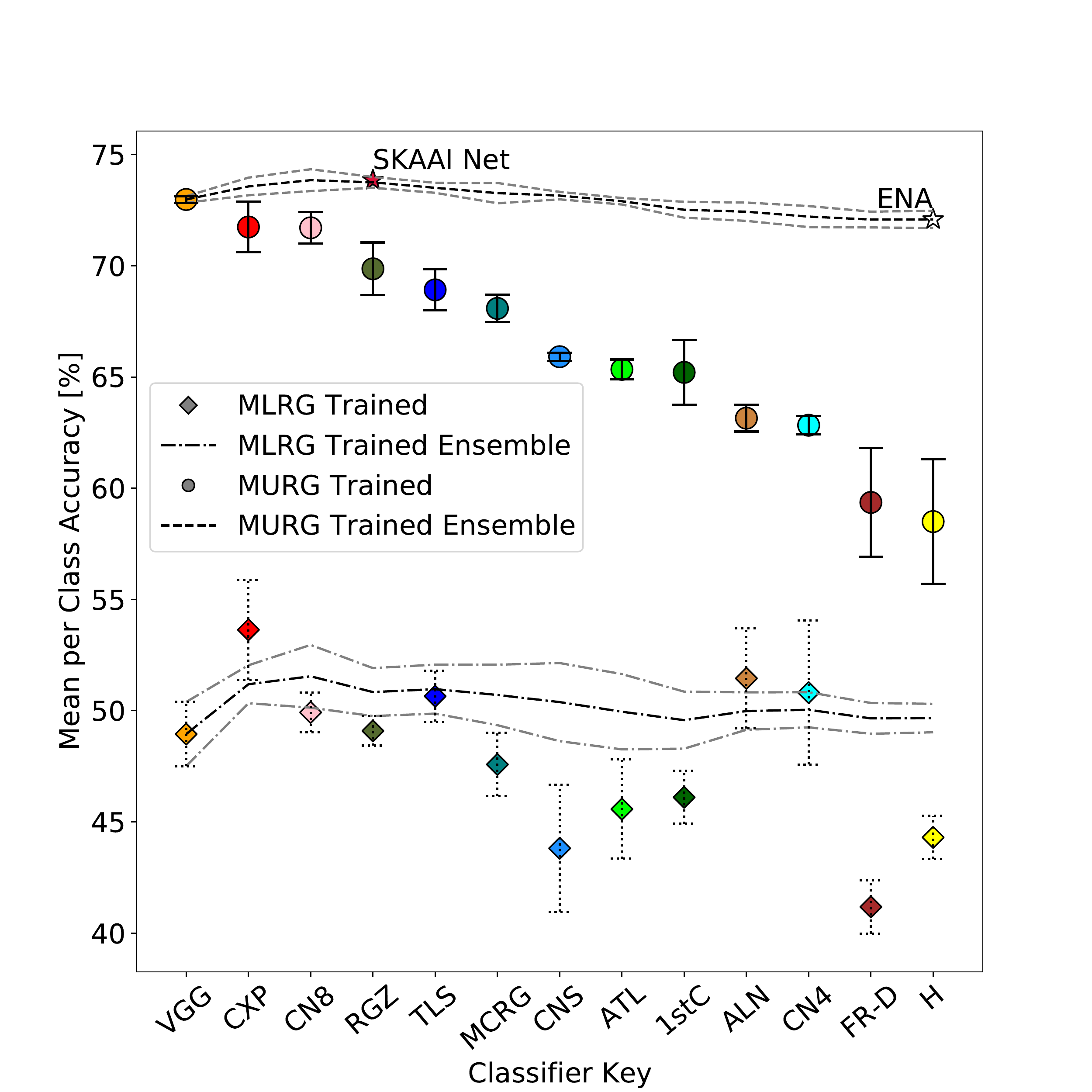}
    \caption{Models trained on a selection from the MURG sample (circles) compared with those trained on the MLRG sample (diamonds) in the ``Overfit'' experiment, giving MPCA averaged over three iterations with different random seed values. The ensemble classifier's averaged MPCA is also given, with the best performing MURG trained ensemble given as SKAAI Net and the ensemble of all classifiers given as ENA. The standard deviation for each architecture and the ensembles are also shown. }
    \label{fig:ensemble}
\end{figure}

In Figure~\ref{fig:ensemble}, we compare the MPCA averaged during three runs of the Overfit experiment (diamond markers) with the MPCA results obtained from the models trained during the MURG Random Split experiment (circle markers). 
It is important to note that the models associated with the two experiments are not tested on the same data (although the intersection between the two datasets is large): the Overfit experiment is tested on the full MURG dataset while the MURG Random Split experiment is tested on a large subset of the MURG dataset (and differs for each run depending on the random seed that was chosen).
All the models associated with the MURG experiment show an increase in recognition performance when they are compared to the models associated with the Overfit experiment (ranging from a 11.7\% to a 24.03\% increase in performance, with an average increase in performance of 18.5\%). The average increase in recognition performance is 3.26 times that of the training data size increase (which increased from 2.2\% to 7.87\% of the total dataset). The result of the top 4 ensemble classifier (SKAAI Net) and the ensemble of all the classifiers (ENA) are also given, the top dashed line represents the results associated with the MURG Random Split ensemble classifier and the bottom dashed line the results associated with Overfit ensemble classifier. Note that the order in this figure is based on the MURG Random Split performance.

The MURG Random Split result is a better indication of architecture performance than the results obtained from the Overfit experiment. The reason being, the architectures are exposed to a larger dataset during training and are, therefore, less prone to overfit. The next section only deals with results obtained from the MURG Random Split experiment.

\section{Results: Computational Performance}
\label{sec:perf_results}

All the results presented in the subsequent subsections were obtained from the MURG Random Split Experiment  (Section~\ref{sec:MURGexperiment}). The results of this experiment is summarized in Table~\ref{tab:table_results_1} and its follow on table.
In Section~\ref{sec:ball_plot}, we report on the MPCA versus the computational complexity of the architectures in Table~\ref{tab:arch_keys}. Note that, for the sake of brevity we sometimes only use the shortened phrase ``architectures'' instead of ``architectures in Table~\ref{tab:arch_keys}'' when referring to the architectures that we considered in this paper. Section~\ref{sec:f1_score_results}, looks at the per class F1-score performance of the architectures (which serves to showcase the trade-off in class performance for each classifier and highlights the shortfalls of a metric such as MPCA). The memory requirements and the classification speeds of the architectures  are discussed in Section~\ref{sec:memory} and Section~\ref{sec:speed}. The receptive field and the effective stride length of the architectures are reported in Section~\ref{sec:rp_esl}. The overall ranking of the architectures is presented in Section~\ref{sec:ranking}. Section~\ref{sec:ensemble_results}, reports on the performance results of the ENA and SKAAI Net (the two ensemble classifiers described in Section~\ref{sec:ensemble}). The data coverage versus the confidence threshold graphs associated with SKAAI Net is presented in Section~\ref{sec:coverage_results}. 

\subsection{Accuracy-rate vs Computational Complexity vs Model Complexity}
\label{sec:ball_plot}

\begin{figure*}
	\includegraphics[width=\textwidth]{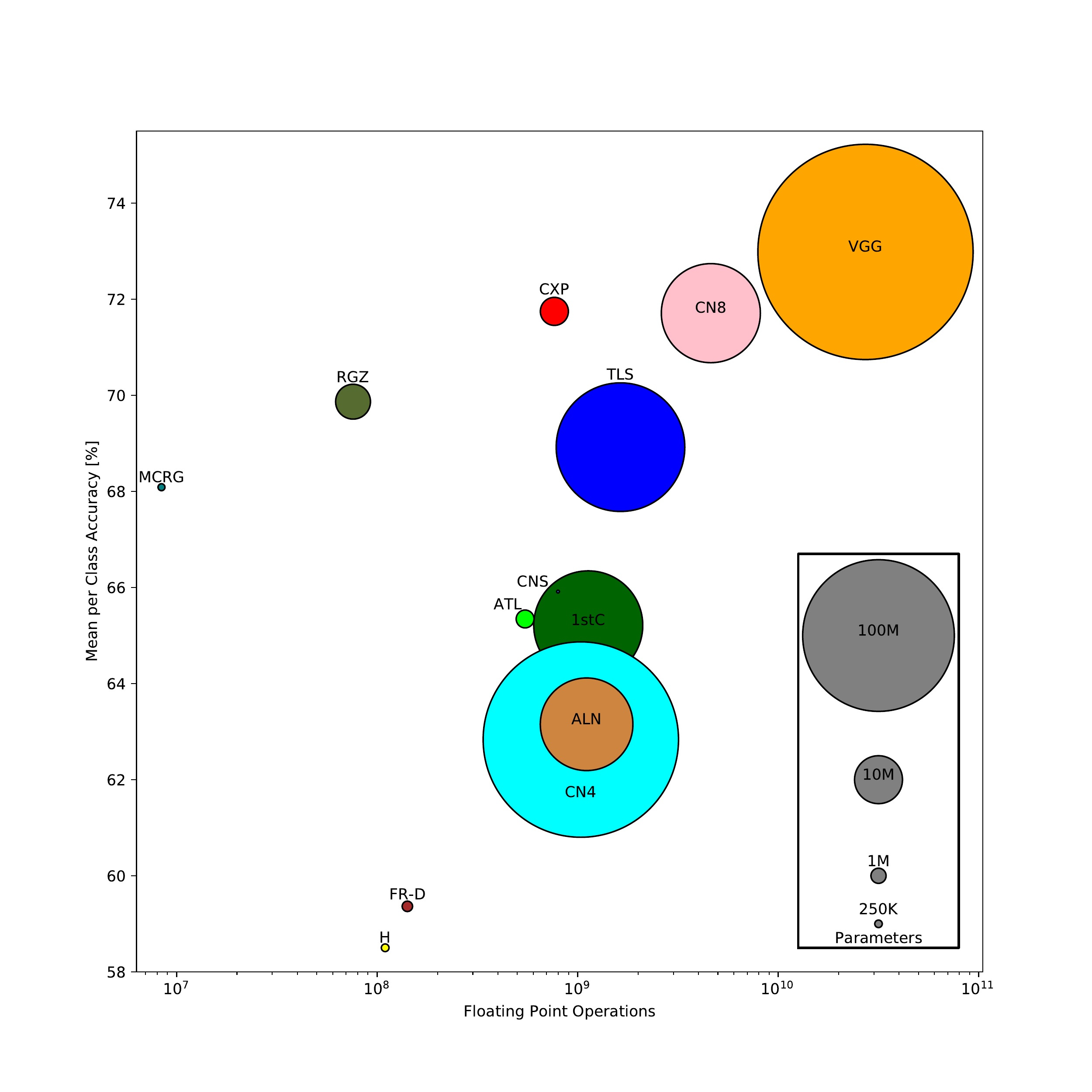}
    \caption{Average accuracy vs Computational Complexity vs Model Complexity: The different CNN architectures are compared based on recognition performance (MPCA on the $y$-axis), computational complexity (FLOPs on the $x$-axis) and model complexity (the number of trainable parameters, given as the circle size). A weak correlation is present between the logarithm of the computational complexity and recognition performance (Pearson correlation coefficient of 0.39).}
    \label{fig:ball_plot}
\end{figure*}

Figure~\ref{fig:ball_plot} reports the MPCA versus the computational complexity of the  architectures in Table~\ref{tab:arch_keys}; for a single forward pass (measured in floating point operations or FLOPs). The size of the markers in Figure~\ref{fig:ball_plot} represent the model complexity of the architectures (measured in the number of trainable parameters). 

The model with the highest MPCA (72.98\%) is {\sc CLARAN} (i.e. VGG16) \citep{Wu2019,Simonyan2014}. The best performing models from the existing literature are {\sc ConvNet8} \citep{Lukic2019b} (71.7\%), {\sc Radio Galaxy Zoo} \citep{Lukic2018} (69.87\%) and {\sc Toothless} \citep{Aniyan2017} (68.92\%). The novel classifier produced for this paper, {\sc ConvXpress}, has the second highest MPCA (71.74\%).

Using classifiers from the general computer vision literature, that perform well on other datasets, as a springboard for architecture development in radio astronomy could potentially save significant computational time. This is evident looking at {\sc Toothless} that was derived from {\sc AlexNet} and and is 5$^{\textrm{th}}$ best classifier in this study, even though this was the first CNN implemented specifically for radio astronomy. 

A very weak correlation between the logarithm of the FLOP count and MPCA is present with a Pearson correlation coefficient of 0.39 \citep{freedman2007}. 

An increase in computational complexity will not necessarily translate into a proportional increase in recognition performance, evidenced by the weak correlation of these two quantities. This is corroborated by the following examples: ATLAS and {\sc Radio Galaxy Zoo} require fewer FLOPS than \citet{Lukic2019b}'s {\sc SimpleNet} and {\sc ConvNet4}, whilst obtaining a better MPCA than the latter two architectures.

The logarithm of the number of parameters and MPCA are even more weakly correlated than the logarithm of the FLOP count and MPCA, with a Pearson correlation coefficient of 0.35. Large models (i.e. higher trainable parameter count) often outperform smaller models in terms of recognition performance, but notable exceptions exist. {\sc ConvXpress} and {\sc Radio Galaxy Zoo} have low parameter counts but have high MPCA scores.

\subsection{Per Class F1-score}
\label{sec:f1_score_results}

\begin{figure*}
	\includegraphics[scale=0.5]{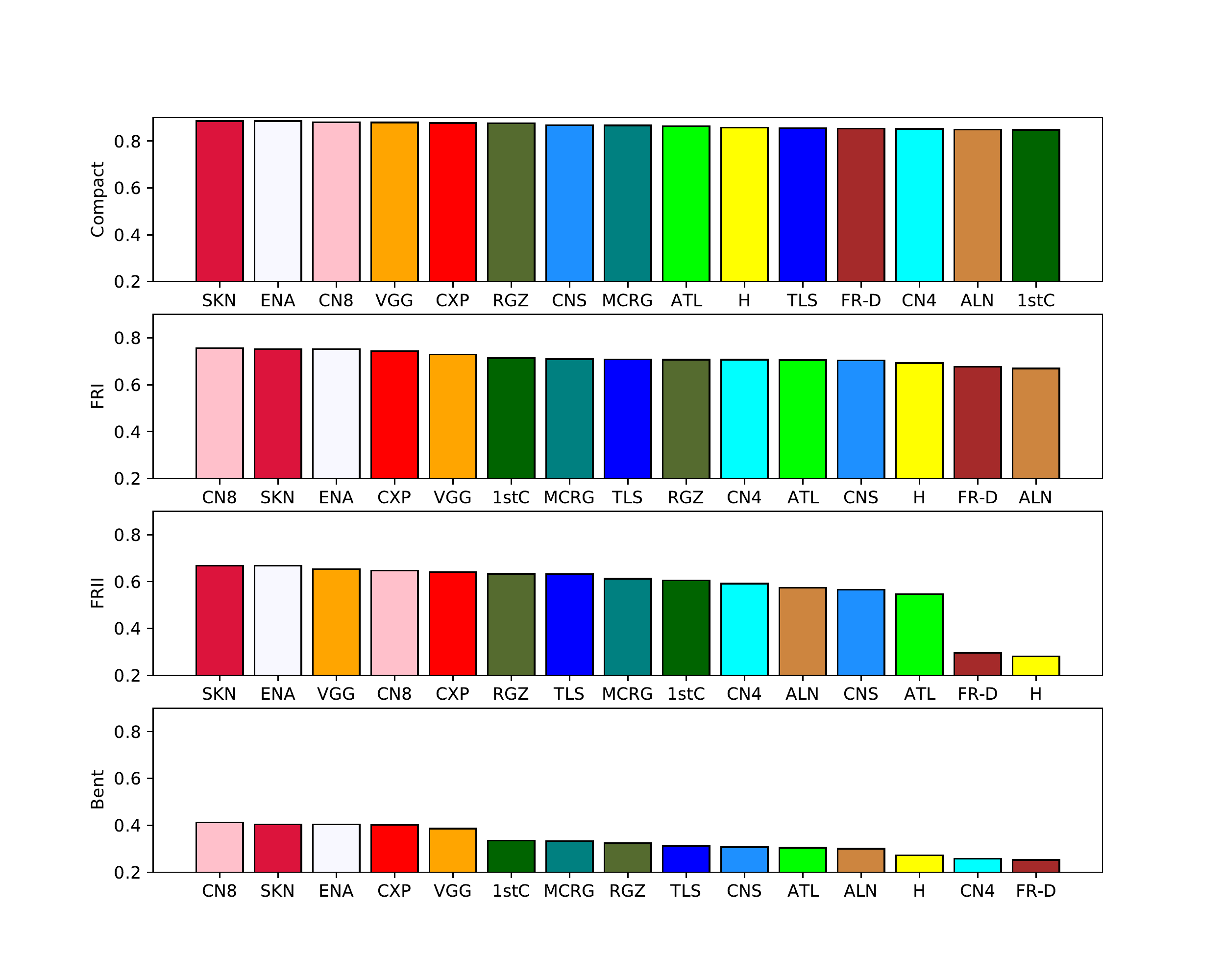}
    \caption{F1-score per class for all architectures. ENA represents an ensemble classifier of all architectures while SKAAI Net (SKN) is an ensemble of the top 4 classifiers.}
    \label{fig:f1_score}
\end{figure*}

Figure~\ref{fig:f1_score}, reports the F1-score of each class sorted by architecture performance. F1-score encapsulates recall and precision into a single metric. More general metrics, such as MPCA and overall accuracy, can be misleading metrics, since a model might score high in either of these metrics by doing exceptional well in one class, whilst underperforming in another class.

Classifiers should in general not be evaluated using a single metric, however, a single metric is sometimes necessary as it can convey information in a concise and succinct manner.

\subsection{Inference Time vs. GPU Memory Usage}
\label{sec:memory}
\begin{figure}
	\includegraphics[scale=0.35]{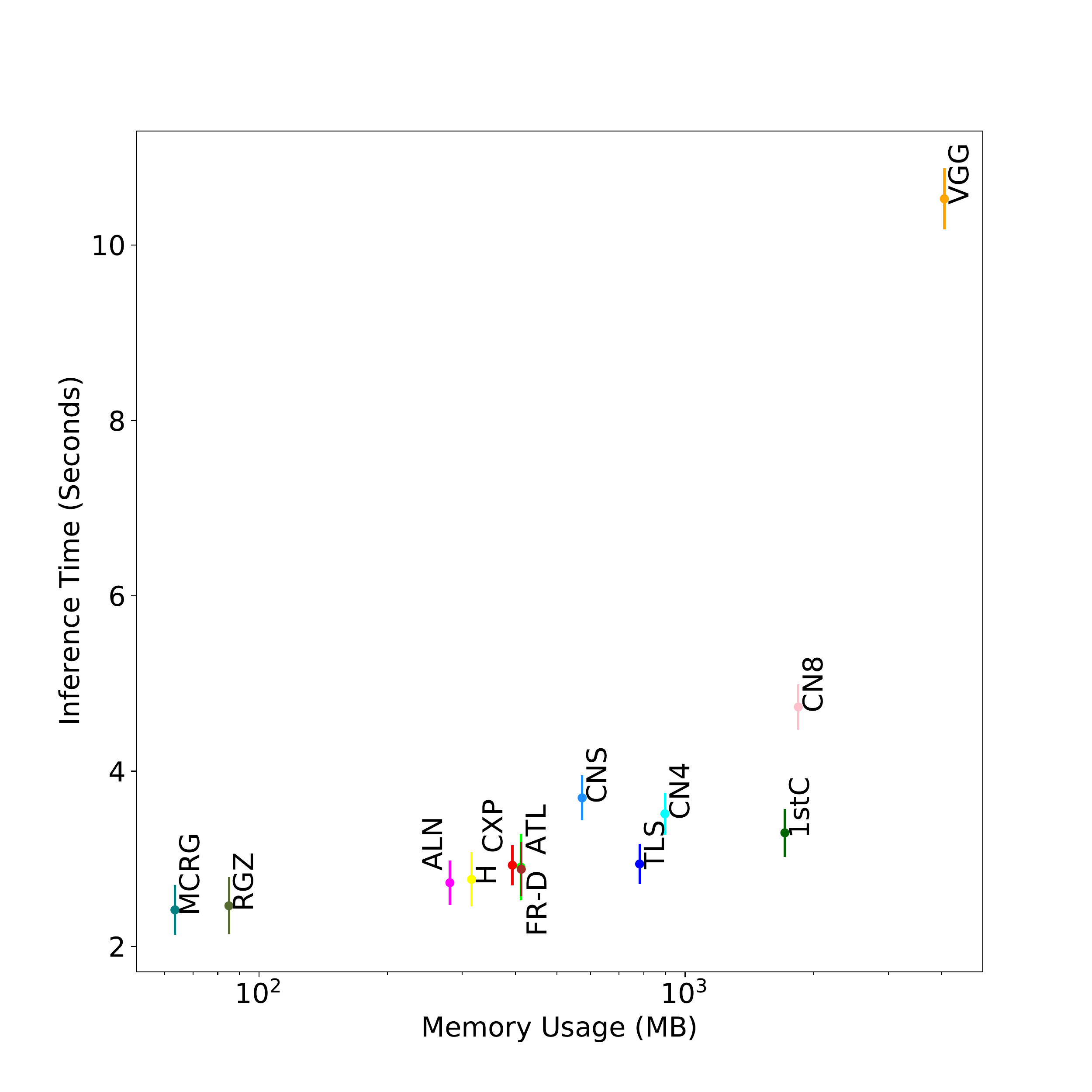}
    \caption{Inference time versus GPU memory usage.}
    \label{fig:memory}
\end{figure}

Figure~\ref{fig:memory}, reports inference time versus theoretical GPU memory usage (at a batch size of 32). As memory usage increases, inference time dramatically increases. The standard deviation associated with the inference time is also depicted in Figure~\ref{fig:memory}. 

\subsection{Classification Speed}
\label{sec:speed}

\begin{figure}
	\includegraphics[scale=0.35]{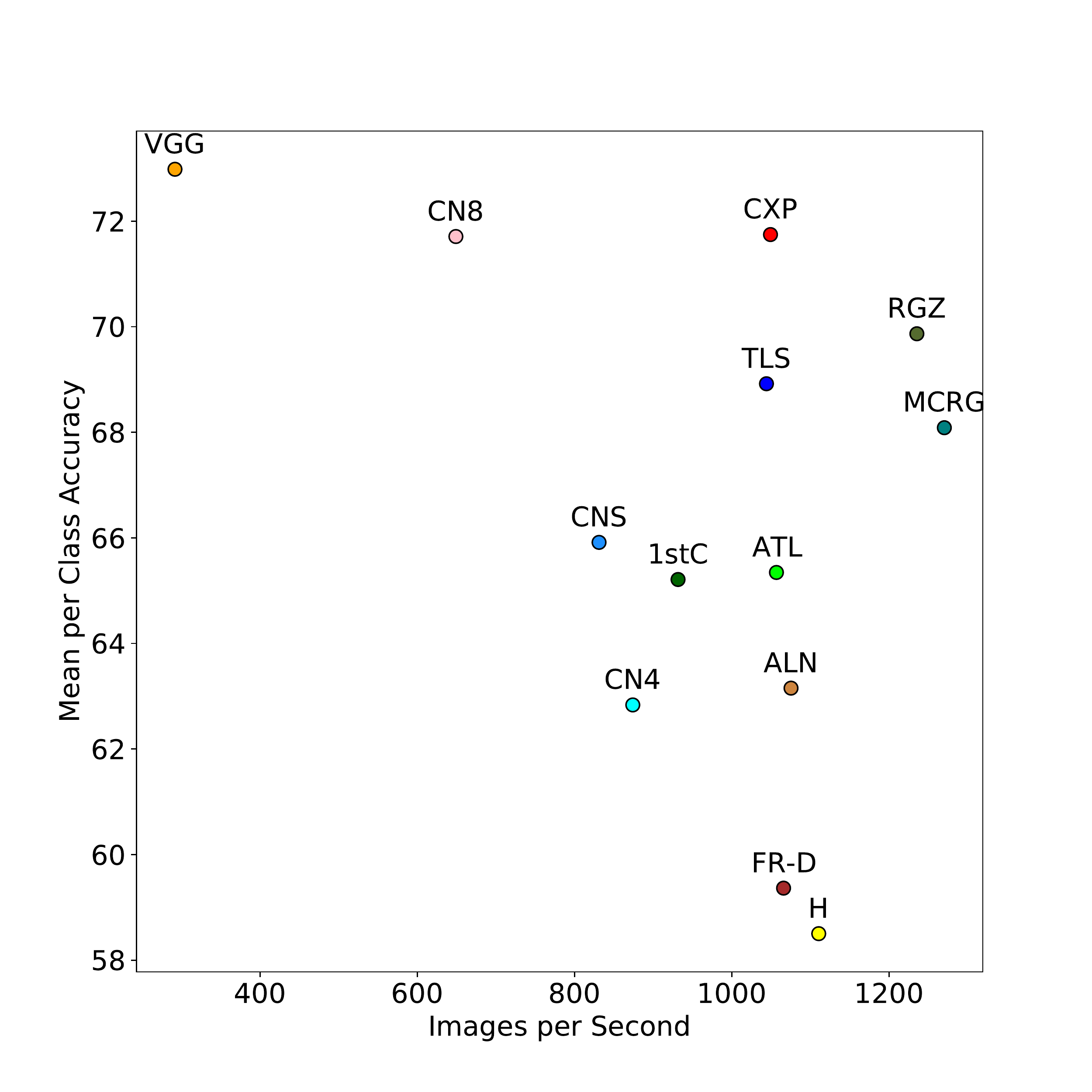}
    \caption{MPCA versus Images per Second: As classification speed increases, recognition performance decreases.}
    \label{fig:speed}
\end{figure}

Classification speed is the number of images an architecture can classify per second at a certain batch size. A batch size of 32 has been used for this experiment. The classification speed of the different architectures are obtained by taking the inverse of the inference times reported in Figure~\ref{fig:memory}. Figure~\ref{fig:speed} reports the MPCA versus classification speed of the different architectures. It is evident from Figure~\ref{fig:speed} that computationally efficient models generally have faster classification speeds. A trade-off, therefore, exists between faster classification and higher recognition performance, at least for standard CNN architectures. Moreover, {\sc MCRGNet} has the fastest classification speed at 1270 images per second with {\sc Radio Galaxy Zoo} close by at 1246 images per second. VGG16 has the slowest classification speed at 291 images per second. 

In comparison, the classification speed of an average person is ``about 250 images in 5 minutes'' or roughly 0.833 images per second \citep{markoff_2012}. The classification task from which this result was obtained is complex, a human classifier had to choose a label from a large number of possibilities, and as such the aforementioned result should be regarded as a lower bound estimate of how fast an average person would be able to classify 250 images.

\subsection{Receptive Field and Effective Stride Length}
\label{sec:rp_esl}

\begin{figure}
	\includegraphics[scale=0.35]{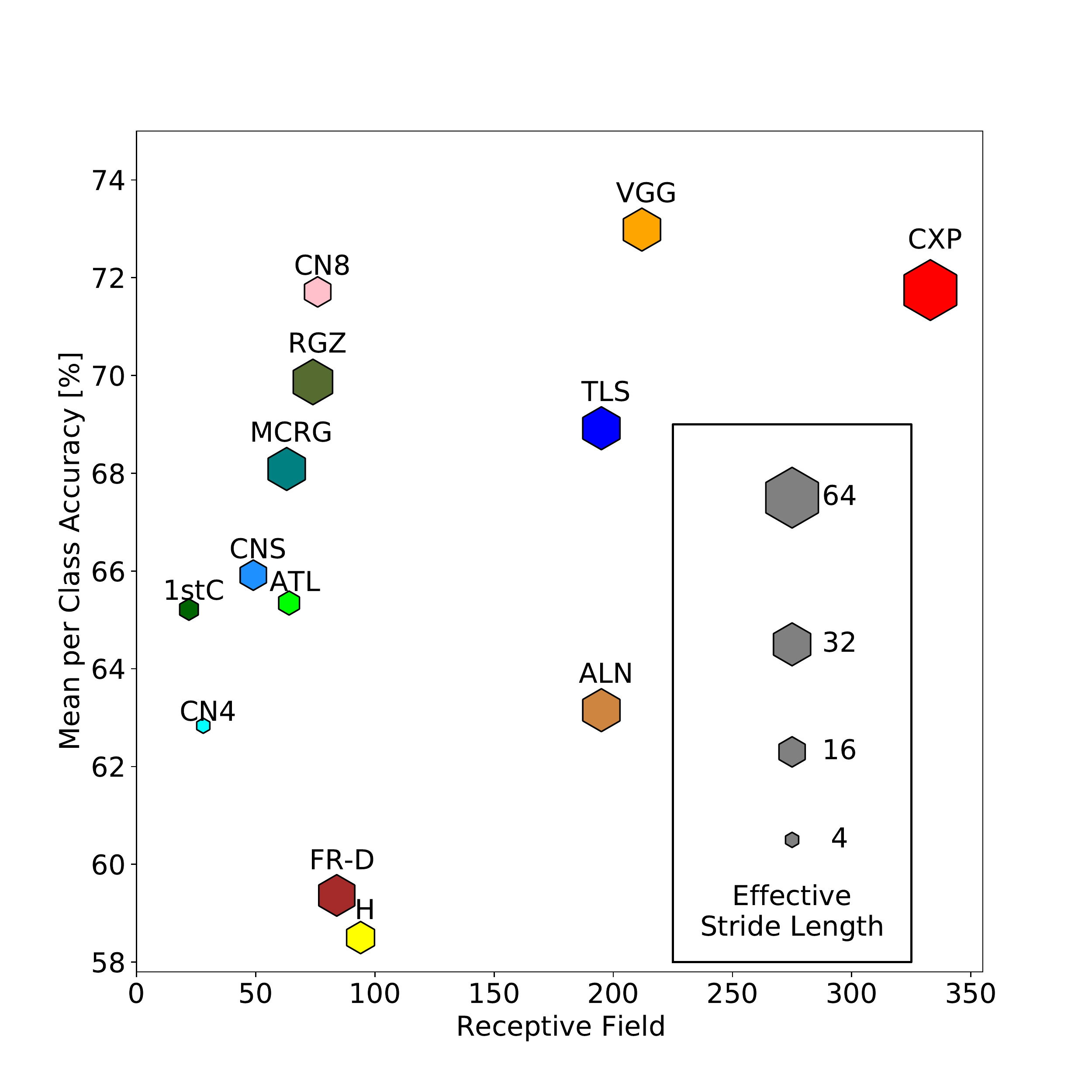}
    \caption{MPCA versus Receptive field size versus effective stride length. In general, as the receptive field and effective stride length increases so does MPCA, care should be taken when designing architectures based only on receptive field size, since a larger receptive field or effective stride length does not always translate into a higher accuracy. Receptive field and effective stride length could, however, serve as a useful metric to better understand the performance of a particular CNN.}
    \label{fig:erf}
\end{figure}

Figure~\ref{fig:erf} shows how receptive field, effective stride length and MPCA relate to one another. The correlation between receptive field and MPCA is very weak (with a Pearson correlation coefficient of 0.431). Effective stride length and MPCA are slightly better correlated (with a Pearson correlation coefficient of 0.436). This is well corroborated by Figure~\ref{fig:erf}. As an example: the {\sc ConvNet8} has a small receptive field and effective stride, but performs comparatively well against architectures that have larger receptive fields and strides than it does. In summary, a larger receptive field and effective stride alone is no guarantee of better classifier performance. 
Using larger strides reduces the number of convolutions applied which results in faster  classification speeds. Applying larger strides in the first layers of the architecture reduces the layer's output size which ultimately decreases the number of FLOPs used by the architecture as it reduces the input sizes of subsequent layers.

\subsection{Ranking}
\label{sec:ranking}
All the architectures listed in Table~\ref{tab:arch_keys} have been ranked in Table~\ref{tab:rank} according to their recognition performance (given as classifier ranking) and their computational performance (given as computational ranking). An overall rank is calculated based on the sum of these two rankings. Please note that this ranking is a relative ranking and that it is limited to the architectures within this study (and the datasets used) and as such cannot be seen as an absolute reflection of architecture standing.

The aforementioned rankings are calculated using a round-robin ``tournament'' in which each architecture is compared to every other architecture (excluding itself) in several different categories. If an architecture achieves a higher or lower score (which depends on the metric of the category under consideration) than a ``competing'' architecture does in a specific category then the former architecture's ranking is incremented by $j$, while the latter architecture's ranking is decremented by $k$. As alluded to before, this comparison is repeated for every category and every architecture-pair. A higher category score is better in the case of recognition performance metric categories, while a lower category score is better in the case of computational requirement metric categories. To establish a classifier ranking, the MPCA and the per class F1-score  of the different architectures are compared with one another (i.e. a total of 5 categories are considered). For the classifier ranking, $k=j=1$. To establish a computational ranking, GPU memory usage, floating point operations and inference time are compared with one another (i.e. a total of 3 categories are considered). For the computational ranking, we also decided upon using $k=j=1$. 

\begin{table}
\caption{Classifier Rankings: the proposed ranking system is based upon recognition performance (classification ranking) and computational performance (computational ranking). {\sc ConvXpress}, {\sc MCRGNet} and {\sc Radio Galaxy Zoo} rank in the top three, each showing a different balance between computational performance and recognition performance. The rankings are calculated using a round-robin ``tournament'' in which each architecture is compared to every other classifier (excluding itself) in each category. MPCA and F1-score for each class are used for the classification ranking while memory usage, FLOP count and inference time is used for the computational ranking.}
\label{tab:rank}
\centering
\setlength{\tabcolsep}{6pt}
\begin{tabular}{lrrr}
\hline
Key  & Classification  & Computational  & \textbf{Overall} \\
& Ranking & Ranking & \textbf{Rank} \\
\hline
CXP  & 46  & 6   & \textbf{52}  \\
MCRG & 14  & 36  & \textbf{50}  \\
RGZ  & 20  & 30  & \textbf{50}  \\
CN8  & 54  & -30 & \textbf{24}  \\
VGG  & 50  & -36 & \textbf{14}  \\
TLS  & 6   & -14 & \textbf{-8}  \\
ATL  & -18 & 8   & \textbf{-10} \\
CNs  & -10 & -10 & \textbf{-20} \\
H    & -42 & 20  & \textbf{-22} \\
1stC & -4  & -18 & \textbf{-22} \\
ALN  & -38 & 12  & \textbf{-26} \\
FR-D & -48 & 10  & \textbf{-38} \\
CN4  & -30 & -14 & \textbf{-44} \\
\hline
\end{tabular}
\end{table}

\subsection{Ensemble Classifier Performance}
\label{sec:ensemble_results}

The recognition performance of SKAAI Net and ENA is given in Figures \ref{fig:ensemble} and \ref{fig:f1_score} (see Section~\ref{sec:ensemble}). Comparing either ensemble's performance with any individual classifier's performance, the ensemble methods outperform individual classifiers in terms of accuracy and outperform most in F1-score. The top 4 classifier (SKAAI Net) performs the best when we consider the MPCA metric (73.85\%) and scores the highest in F1-score in 2 classes, being second in the FRI and Bent classes (indicating that even though MPCA has its shortcomings it remains a helpful keystone metric to use for evaluating architecture performance). The ensemble of all the models (ENA), on the other hand, has an MPCA of 72.08\%, just slightly below the highest MPCA of the single classifiers ({\sc ConvXpress}, 72.98\%) and performs well in F1-score. 

\begin{figure}
	\includegraphics[scale=0.3]{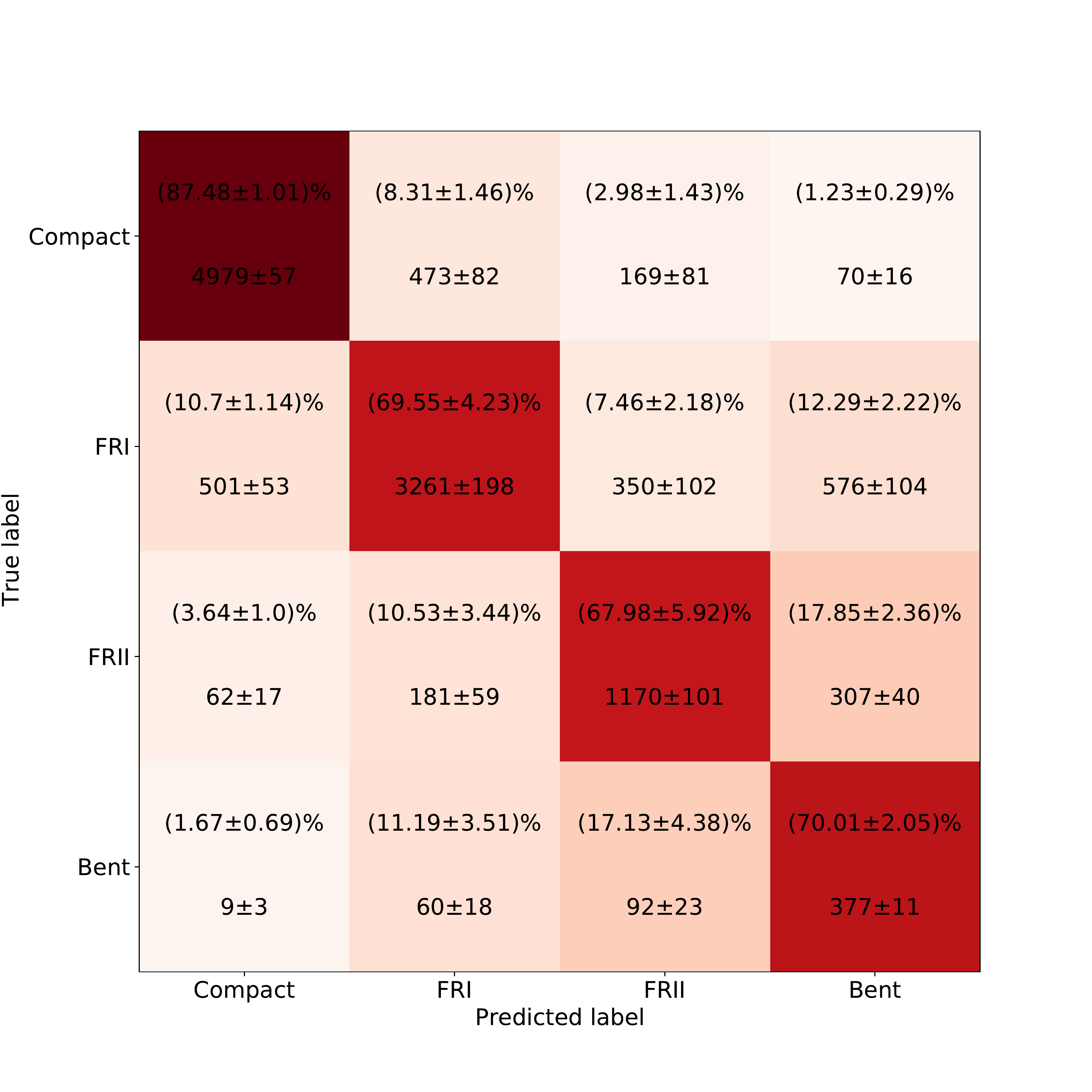}
    \caption{Confusion Matrix of the SKAAI Net ensemble, averaged over three runs. The normalized confusion matrix is given in percentage in each square, above the number of sources classified }
    \label{fig:confusion_matrix}
\end{figure}

SKAAI Net's confusion matrix is depicted in Figure~\ref{fig:confusion_matrix}. We only provide SKAAI Net's confusion matrix here as it outperforms ENA. The main diagonal of the matrix shows the number of correctly classified images, with the columns indicating the classifier's prediction and the rows the actual label of each image. The percentages are the normalized values for each class. Percentage wise, the most misclassifications are FRII that are being labelled as bent-tails (17.85\%), but in absolute terms more FRIs are misclassified as bent-tails on average (576).

Overall, SKAAI Net provides an ensemble classifier that reduces classifier specific shortcomings in regards to recognition performance.
It should be noted that these ensemble methods will require a significant amount of computational resources in order to run (as they consist of more than one model), resulting in a much slower classification speed. 

ENA in particular has an exceptionally large computational footprint, as it is made up of all the classifiers in this study, which makes it infeasible for deployment in production (it requires \textasciitilde{11.8GB} of GPU memory). 
Whether the marginal gains SKAAI Net and ENA make in MPCA and F1-score is worth the significant increase in computational requirements that are needed to achieve those gains is, therefore, highly debatable.

\subsection{Dataset Coverage and Classifier Confidence}
\label{sec:coverage_results}

\begin{figure*}
	\includegraphics[width=\textwidth]{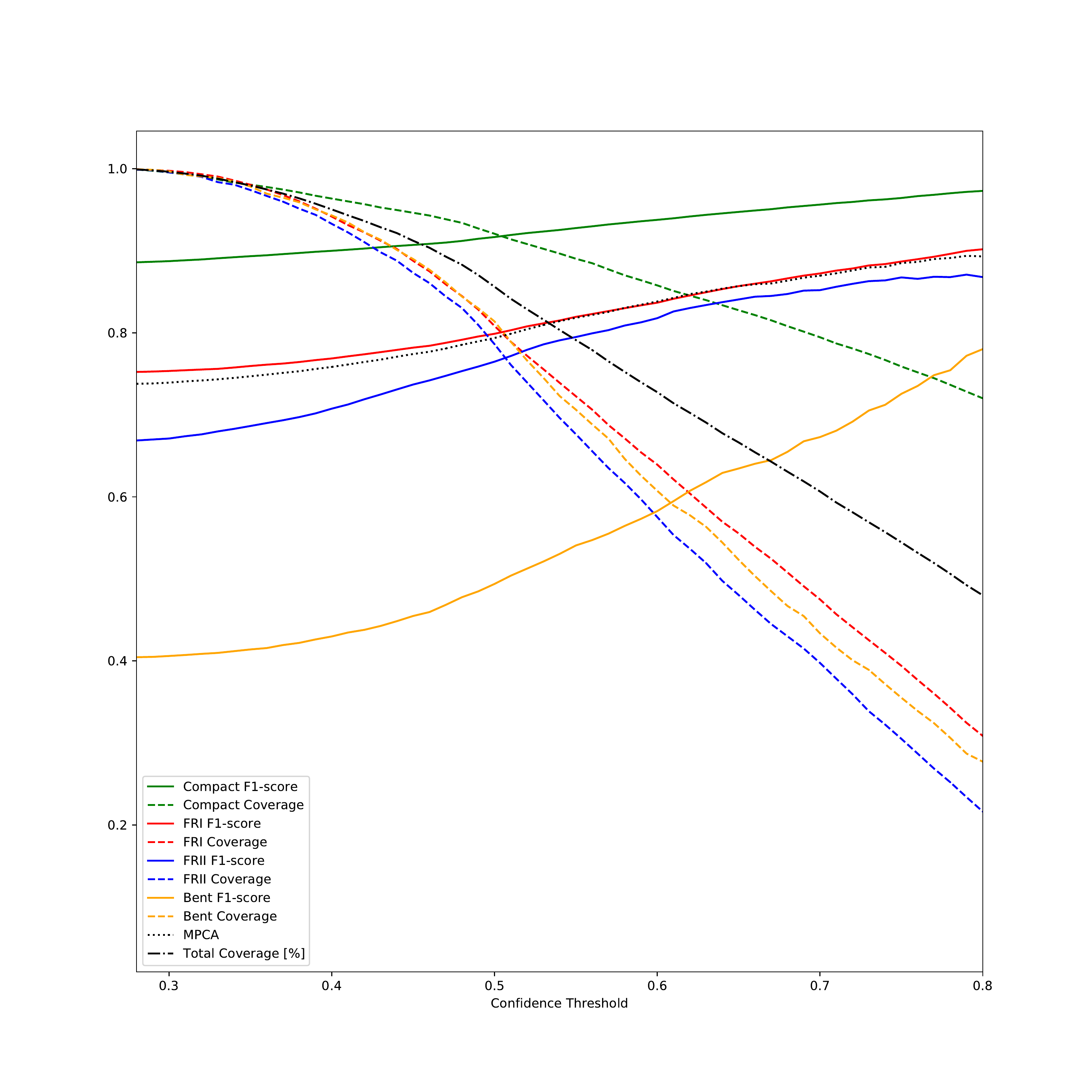}
    \caption{SKAAI Net Confidence Threshold: Dataset Coverage vs Recognition Performance.}
    \label{fig:coverage}
\end{figure*}

Figure~\ref{fig:coverage}, reports on the percentage dataset coverage per class and F1-score per class of SKAAI Net at different confidence thresholds. It also reports the percentage coverage for the entire dataset and MPCA.

Compact sources have the highest F1-score overall and the smallest decrease in coverage. Bent tails sees the largest increase in F1-score. FRII coverage drops at a faster rate than the other classes. 

Knowing the data coverage behaviour of a classifier is important. It allows an estimate of the resources that would be required if this were to be incorporated with subject-matter experts into the classification pipeline, i.e. on average exactly how many images would be thrown out by the classification system at a specific certainty threshold, which in turn would help us estimate the number of man hours that would be required to manually classify the sources that were thrown out. On the other hand, if the human resource availability is known, the certainty threshold can be adapted.

\section{Conclusions}

Two experiments were performed in this study. The first experiment assessed overfitting on the MLRG dataset, that has large intersections with the training sets used in most studies in Table~\ref{tab:arch_keys} (see Section~\ref{sec:experiments} and Section~\ref{sec:overfit_results}). The second experiment analysed the computational cost of existing CNN architectures used for radio galaxy morphological classification (see Section~\ref{sec:perf_results}). The results from these experiment suggests that when evaluating an architecture's performance careful attention should be paid to the size of the training set being used, otherwise the results obtained could potentially not be a true reflection of architecture performance (see Figure~\ref{fig:overfit} and Figure~\ref{fig:ensemble}). Furthermore, there exists a trade-off between recognition performance and computational cost. These two factors should be carefully weighed up against each other when deciding on which architecture to use in production. In addition to considering recognition performance metrics like MPCA and F1-score, one should also consider computational cost metrics like memory usage, floating point operations used and classification speed. From all of these metrics a ``best'' architecture can be chosen for deployment, based on the computational resources that are available. 

There are also a few other minor conclusions that can be drawn from the results obtained from the experiments we conducted in this paper:

\begin{description}
\item[\textbf{Architecture}] A few design choices for CNN architectures can speed up model performance while driving down resource costs. A larger kernel's receptive field is equivalent to several smaller receptive fields when these are stacked without a pooling layer in between with the added bonus that more layers of non-linearity are added through more ReLU activations while driving down the number of parameters. This was originally used within the VGG architectue family developed by \citet{Simonyan2014}. Several of the architectures in Table~\ref{tab:arch_keys} build on these design decisions, specifically ConvNet-8 \citep{Lukic2019a}.

\item[\textbf{{\sc ConvXpress}}] {\sc ConvXpress} utilizes the aforementioned stacking strategy. It also uses a non-standard stride length which indirectly reduces its computational cost. This architecture performs well when compared to the other architectures in Table~\ref{tab:arch_keys} (see Table~\ref{tab:rank}).  

\item[\textbf{Parameters}] Since model complexity is weakly correlated with recognition performance, an increase in parameters is, therefore, likely to translate into an increase in recognition performance (this is however not guaranteed). Increasing your recognition performance by simply utilizing more and more trainable parameters is discouraged as it is a strategy that can lead to overfitting (it also does not scale well). Furthermore, an increase in trainable parameters will increase computational complexity, training time and GPU memory usage.
Overall Deep Learning models are viewed as inefficient in exploiting their full learning power \citep{Muhammed2017} given the large number of parameters they require relative to other machine learning approaches.

\item[\textbf{FLOPs}] Computational complexity (given as FLOPs) and recognition performance (approximated as MPCA) are weakly correlated (see Figure~\ref{fig:ball_plot}). Utilizing more computational resources is, therefore, likely to result in at least marginal increases in recognition performance. As we have hinted at previously, increasing your recognition performance by simply utilizing more and more computational resources is frowned upon as it is a strategy that does not scale well.

\item[\textbf{Classification Speed}]
Generally, models with a higher MPCA, classify slower than those with a lower MPCA (see Figure~\ref{fig:speed}). The trade-off between classification speed and recognition performance is evident as model MPCA decreases with an increase in images classified per second.

\item[\textbf{Receptive field and Stride length}] A large receptive field and stride length does not guarantee good recognition performance. These two metrics can, however, help explain the performance of a particular architecture.  
\item[\textbf{Ranking}] CNNs can be ranked according to their recognition performance results and the computational resources that they require. The ranking we obtained doing just this is presented in Table~\ref{tab:rank}. It is, however, important to realize that this study is not exhaustive enough to provide us with an absolute ranking of the architectures in Table~\ref{tab:arch_keys}. A more extensive study that considers all possible combinations of hyperparameters would be required for us to achieve the aforementioned goal. Such a study would, however, be computationally infeasible. This study does, however, provide us with a useful pragmatic ranking as the hyperparameters were chosen in accordance with excepted guidelines. 
\item{\textbf{Ensemble}} While the ensemble methods do produce better results, the significant increase in computational requirements associated with using them is not proportional to the gain in recognition performance that using them offers. An option left unexplored in this study is the creation of either a tree classifier (such as was done in the original {\sc MCRGNet} study \citep{Ma2019}) or a fusion classifier with a voting scheme (as used by {\sc Toothless} \citep{Aniyan2017}). Both of these approaches requires the training of several models that specialize in the classification of only two classes. A majority vote of a single class is indicative of a high confidence of such a prediction, while a mixed vote indicates uncertainty (such a source would be marked for inspection by a subject-matter expert).
\item{\textbf{Coverage}} Data coverage analysis results can be used to integrate subject-matter experts into a classification pipeline (or at the very least the ability to flag sources that the classifier is uncertain of can be added) (see Section~\ref{sec:coverage_results}). Which confidence threshold is best suited for this endeavour is not explored, since this will rely on the availability of the following: subject-matter experts and computational resources.
\end{description}
For an in-depth comparison of the more recent architectures that are being used for image recognition, please refer to the work done by \citet{Muhammed2017}.
Training CNNs for the purpose of image classification and specifically radio galaxy classification, has become a relatively easy task to set up given the computational resources available at present. But as Jitendra Malik~\footnote{Arthur J. Chick Professor of Electrical Engineering and Computer Sciences at the University of California, Berkeley}, one of the seminal figures in computer vision, stated ``There are many problems in [computer] vision where getting 50 percent of the solution you can get in one minute, getting to 90 percent can take you a day, getting to 99 percent may take you five years and 99.99 percent, may not happen in your lifetime'' \citep{fridman_malik_2020}.

The lack of large sets of annotated training data remains one of the greatest challenges in assessing and improving the general recognition performance of CNNs (and all image classification algorithms'). In addition, determining the computational resources a model requires is important to consider when selecting an architecture for deployment. The framework and experiments laid out in this study will hopefully be able to help shape the future of image recognition development. 

\section*{Acknowledgements}
MV and MP acknowledge financial support from the Inter-University Institute for Data Intensive Astronomy (IDIA). IDIA is a partnership of the University of Cape Town, the University of Pretoria and the University of the Western Cape. We acknowledge the use of the ilifu cloud computing facility - www.ilifu.ac.za, a partnership between the University of Cape Town, the University of the Western Cape, the University of Stellenbosch, Sol Plaatje University, the Cape Peninsula University of Technology and the South African Radio Astronomy Observatory. The ilifu facility is supported by contributions from the Inter-University Institute for Data Intensive Astronomy (IDIA - a partnership between the University of Cape Town, the University of Pretoria and the University of the Western Cape), the Computational Biology division at UCT and the Data Intensive Research Initiative of South Africa (DIRISA). This work is based on the research supported wholly/in part by the National Research Foundation of South Africa (Grant Numbers 117275/119488/121291).

\section*{Data Availability Statement}
The data underlying this article will be shared on reasonable request to the corresponding author.
%
%
%
%
\bibliographystyle{mnras}
\bibliography{main} 

\begin{thebibliography}{}
\makeatletter
\relax
\def\mn@urlcharsother{\let\do\@makeother \do\$\do\&\do\#\do\^\do\_\do\%\do\~}
\def\mn@doi{\begingroup\mn@urlcharsother \@ifnextchar [ {\mn@doi@}
  {\mn@doi@[]}}
\def\mn@doi@[#1]#2{\def\@tempa{#1}\ifx\@tempa\@empty \href
  {http://dx.doi.org/#2} {doi:#2}\else \href {http://dx.doi.org/#2} {#1}\fi
  \endgroup}
\def\mn@eprint#1#2{\mn@eprint@#1:#2::\@nil}
\def\mn@eprint@arXiv#1{\href {http://arxiv.org/abs/#1} {{\tt arXiv:#1}}}
\def\mn@eprint@dblp#1{\href {http://dblp.uni-trier.de/rec/bibtex/#1.xml}
  {dblp:#1}}
\def\mn@eprint@#1:#2:#3:#4\@nil{\def\@tempa {#1}\def\@tempb {#2}\def\@tempc
  {#3}\ifx \@tempc \@empty \let \@tempc \@tempb \let \@tempb \@tempa \fi \ifx
  \@tempb \@empty \def\@tempb {arXiv}\fi \@ifundefined
  {mn@eprint@\@tempb}{\@tempb:\@tempc}{\expandafter \expandafter \csname
  mn@eprint@\@tempb\endcsname \expandafter{\@tempc}}}

\bibitem[\protect\citeauthoryear{{Alger} et~al.,}{{Alger}
  et~al.}{2018}]{Alger2018}
{Alger} M.~J.,  et~al., 2018, \mn@doi [\mnras] {10.1093/mnras/sty1308}, \href
  {https://ui.adsabs.harvard.edu/abs/2018MNRAS.478.5547A} {478, 5547}

\bibitem[\protect\citeauthoryear{{Alhassan}, {Taylor}  \& {Vaccari}}{{Alhassan}
  et~al.}{2018}]{Alhassan2018}
{Alhassan} W.,  {Taylor} A.~R.,   {Vaccari} M.,  2018, \mn@doi [\mnras]
  {10.1093/mnras/sty2038}, \href
  {https://ui.adsabs.harvard.edu/abs/2018MNRAS.480.2085A} {480, 2085}

\bibitem[\protect\citeauthoryear{{Aniyan} \& {Thorat}}{{Aniyan} \&
  {Thorat}}{2017}]{Aniyan2017}
{Aniyan} A.~K.,  {Thorat} K.,  2017, \mn@doi [\apjs]
  {10.3847/1538-4365/aa7333}, \href
  {https://ui.adsabs.harvard.edu/abs/2017ApJS..230...20A} {230, 20}

\bibitem[\protect\citeauthoryear{Araujo, Norris  \& Sim}{Araujo
  et~al.}{2019}]{araujo2019}
Araujo A.,  Norris W.,   Sim J.,  2019, Distill, 4, e21

\bibitem[\protect\citeauthoryear{{Baldi}, {Capetti}  \& {Massaro}}{{Baldi}
  et~al.}{2018}]{Baldi2018}
{Baldi} R.~D.,  {Capetti} A.,   {Massaro} F.,  2018, \mn@doi [\aap]
  {10.1051/0004-6361/201731333}, \href
  {https://ui.adsabs.harvard.edu/abs/2018A&A...609A...1B} {609, A1}

\bibitem[\protect\citeauthoryear{{Banfield} et~al.,}{{Banfield}
  et~al.}{2015}]{Banfield2015}
{Banfield} J.~K.,  et~al., 2015, \mn@doi [\mnras] {10.1093/mnras/stv1688},
  \href {https://ui.adsabs.harvard.edu/abs/2015MNRAS.453.2326B} {453, 2326}

\bibitem[\protect\citeauthoryear{Becker, White  \& Helfand}{Becker
  et~al.}{1995}]{becker1995}
Becker R.~H.,  White R.~L.,   Helfand D.~J.,  1995, \apj, 450, 559

\bibitem[\protect\citeauthoryear{{Best} \& {Heckman}}{{Best} \&
  {Heckman}}{2012}]{Best2012}
{Best} P.~N.,  {Heckman} T.~M.,  2012, \mn@doi [\mnras]
  {10.1111/j.1365-2966.2012.20414.x}, \href
  {https://ui.adsabs.harvard.edu/abs/2012MNRAS.421.1569B} {421, 1569}

\bibitem[\protect\citeauthoryear{{Braun}, {Bourke}, {Green}, {Keane}  \&
  {Wagg}}{{Braun} et~al.}{2015}]{SKA2015}
{Braun} R.,  {Bourke} T.,  {Green} J.~A.,  {Keane} E.,   {Wagg} J.,  2015, in
  Advancing Astrophysics with the Square Kilometre Array (AASKA14). p.~174

\bibitem[\protect\citeauthoryear{{Braun}, {Bonaldi}, {Bourke}, {Keane}  \&
  {Wagg}}{{Braun} et~al.}{2019}]{SKA2019}
{Braun} R.,  {Bonaldi} A.,  {Bourke} T.,  {Keane} E.,   {Wagg} J.,  2019, arXiv
  e-prints, \href {https://ui.adsabs.harvard.edu/abs/2019arXiv191212699B} {p.
  arXiv:1912.12699}

\bibitem[\protect\citeauthoryear{{Capetti}, {Massaro}  \& {Baldi}}{{Capetti}
  et~al.}{2017a}]{Capetti2017a}
{Capetti} A.,  {Massaro} F.,   {Baldi} R.~D.,  2017a, \mn@doi [\aap]
  {10.1051/0004-6361/201629287}, \href
  {https://ui.adsabs.harvard.edu/abs/2017A&A...598A..49C} {598, A49}

\bibitem[\protect\citeauthoryear{{Capetti}, {Massaro}  \& {Baldi}}{{Capetti}
  et~al.}{2017b}]{Capetti2017b}
{Capetti} A.,  {Massaro} F.,   {Baldi} R.~D.,  2017b, \mn@doi [\aap]
  {10.1051/0004-6361/201630247}, \href
  {https://ui.adsabs.harvard.edu/abs/2017A&A...601A..81C} {601, A81}

\bibitem[\protect\citeauthoryear{{Cheung}}{{Cheung}}{2007}]{Cheung2007}
{Cheung} C.~C.,  2007, \mn@doi [\aj] {10.1086/513095}, \href
  {https://ui.adsabs.harvard.edu/abs/2007AJ....133.2097C} {133, 2097}

\bibitem[\protect\citeauthoryear{Chollet et~al.}{Chollet
  et~al.}{2015}]{chollet2015}
Chollet F.,  et~al., 2015, Keras, \url{https://keras.io}

\bibitem[\protect\citeauthoryear{Cire{\c{s}}an, Meier, Gambardella  \&
  Schmidhuber}{Cire{\c{s}}an et~al.}{2010}]{cirecsan2010}
Cire{\c{s}}an D.~C.,  Meier U.,  Gambardella L.~M.,   Schmidhuber J.,  2010,
  Neural Comp., 22, 3207

\bibitem[\protect\citeauthoryear{{Cotton} et~al.,}{{Cotton}
  et~al.}{2020}]{Cotton2020a}
{Cotton} W.~D.,  et~al., 2020, \mn@doi [\mnras] {10.1093/mnras/staa1240}, \href
  {https://ui.adsabs.harvard.edu/abs/2020MNRAS.495.1271C} {495, 1271}

\bibitem[\protect\citeauthoryear{Deng, Dong, Socher, Li, Li  \& Fei-Fei}{Deng
  et~al.}{2009}]{deng2009}
Deng J.,  Dong W.,  Socher R.,  Li L.-J.,  Li K.,   Fei-Fei L.,  2009. IEEE
  Conference on Computer Vision and Pattern Recognition.
Imagenet: A large-scale hierarchical image database, Miami, Florida, p.~248

\bibitem[\protect\citeauthoryear{{Dieleman}, {Willett}  \& {Dambre}}{{Dieleman}
  et~al.}{2015}]{Dieleman2015}
{Dieleman} S.,  {Willett} K.~W.,   {Dambre} J.,  2015, \mn@doi [\mnras]
  {10.1093/mnras/stv632}, \href
  {https://ui.adsabs.harvard.edu/abs/2015MNRAS.450.1441D} {450, 1441}

\bibitem[\protect\citeauthoryear{{Ekers}, {Fanti}, {Lari}  \& {Parma}}{{Ekers}
  et~al.}{1978}]{Ekers1978}
{Ekers} R.~D.,  {Fanti} R.,  {Lari} C.,   {Parma} P.,  1978, \mn@doi [\nat]
  {10.1038/276588a0}, \href
  {https://ui.adsabs.harvard.edu/abs/1978Natur.276..588E} {276, 588}

\bibitem[\protect\citeauthoryear{{Elmegreen} \& {Elmegreen}}{{Elmegreen} \&
  {Elmegreen}}{1987}]{Elmegreen1987}
{Elmegreen} D.~M.,  {Elmegreen} B.~G.,  1987, \mn@doi [\apj] {10.1086/165034},
  \href {https://ui.adsabs.harvard.edu/abs/1987ApJ...314....3E} {314, 3}

\bibitem[\protect\citeauthoryear{{Fanaroff} \& {Riley}}{{Fanaroff} \&
  {Riley}}{1974}]{Fanaroff1974}
{Fanaroff} B.~L.,  {Riley} J.~M.,  1974, \mn@doi [\mnras]
  {10.1093/mnras/167.1.31P}, \href
  {https://ui.adsabs.harvard.edu/abs/1974MNRAS.167P..31F} {167, 31\textsc{P}}

\bibitem[\protect\citeauthoryear{Freedman, Pisani  \& Purves}{Freedman
  et~al.}{2007}]{freedman2007}
Freedman D.,  Pisani R.,   Purves R.,  2007, WW Norton \& Company, New York

\bibitem[\protect\citeauthoryear{Fridman \& Malik}{Fridman \&
  Malik}{2020}]{fridman_malik_2020}
Fridman L.,  Malik J.,  2020, 110 – Jitendra Malik: Computer Vision

\bibitem[\protect\citeauthoryear{Fukushima}{Fukushima}{1980}]{Fukushima1980}
Fukushima K.,  1980, Biol. Cybernetics, 36, 193

\bibitem[\protect\citeauthoryear{{Garofalo} \& {Singh}}{{Garofalo} \&
  {Singh}}{2019}]{Garofalo2019}
{Garofalo} D.,  {Singh} C.~B.,  2019, \mn@doi [\apj]
  {10.3847/1538-4357/aaf056}, \href
  {https://ui.adsabs.harvard.edu/abs/2019ApJ...871..259G} {871, 259}

\bibitem[\protect\citeauthoryear{{Gendre} \& {Wall}}{{Gendre} \&
  {Wall}}{2008}]{Gendre2008}
{Gendre} M.~A.,  {Wall} J.~V.,  2008, \mn@doi [\mnras]
  {10.1111/j.1365-2966.2008.13792.x}, \href
  {https://ui.adsabs.harvard.edu/abs/2008MNRAS.390..819G} {390, 819}

\bibitem[\protect\citeauthoryear{{Gendre}, {Best}  \& {Wall}}{{Gendre}
  et~al.}{2010}]{Gendre2010}
{Gendre} M.~A.,  {Best} P.~N.,   {Wall} J.~V.,  2010, \mn@doi [\mnras]
  {10.1111/j.1365-2966.2010.16413.x}, \href
  {https://ui.adsabs.harvard.edu/abs/2010MNRAS.404.1719G} {404, 1719}

\bibitem[\protect\citeauthoryear{Gheller, Vazza  \& Bonafede}{Gheller
  et~al.}{2018}]{Gheller2018}
Gheller C.,  Vazza F.,   Bonafede A.,  2018, \mn@doi [\mnras]
  {10.1093/mnras/sty2102}, \href {http://dx.doi.org/10.1093/mnras/sty2102}
  {480, 3749}

\bibitem[\protect\citeauthoryear{{Gopal-Krishna} \& {Wiita}}{{Gopal-Krishna} \&
  {Wiita}}{2000}]{Gopal2000}
{Gopal-Krishna} {Wiita} P.~J.,  2000, \aap, \href
  {https://ui.adsabs.harvard.edu/abs/2000A&A...363..507G} {363, 507}

\bibitem[\protect\citeauthoryear{{Harwood}, {Vernstrom}  \& {Stroe}}{{Harwood}
  et~al.}{2020}]{Harwood2020}
{Harwood} J.~J.,  {Vernstrom} T.,   {Stroe} A.,  2020, \mn@doi [\mnras]
  {10.1093/mnras/stz3069}, \href
  {https://ui.adsabs.harvard.edu/abs/2020MNRAS.491..803H} {491, 803}

\bibitem[\protect\citeauthoryear{{Hine} \& {Longair}}{{Hine} \&
  {Longair}}{1979}]{Hine1979}
{Hine} R.~G.,  {Longair} M.~S.,  1979, \mn@doi [\mnras]
  {10.1093/mnras/188.1.111}, \href
  {https://ui.adsabs.harvard.edu/abs/1979MNRAS.188..111H} {188, 111}

\bibitem[\protect\citeauthoryear{{Hosenie}}{{Hosenie}}{2018}]{Hosenie2018}
{Hosenie} Z.~B.,  2018, Master's thesis, North-West Univ., Potchefstroom

\bibitem[\protect\citeauthoryear{{Hubble}}{{Hubble}}{1926}]{Hubble1926}
{Hubble} E.~P.,  1926, \mn@doi [\apj] {10.1086/143018}, \href
  {https://ui.adsabs.harvard.edu/abs/1926ApJ....64..321H} {64, 321}

\bibitem[\protect\citeauthoryear{Hubel \& Wiesel}{Hubel \&
  Wiesel}{1968}]{Hubel1968}
Hubel D.~H.,  Wiesel T.~N.,  1968, The J. of Phys., 195, 215

\bibitem[\protect\citeauthoryear{Kelley}{Kelley}{1960}]{kelley1960}
Kelley H.~J.,  1960, ARS J., 30, 947

\bibitem[\protect\citeauthoryear{{Kozie{\l}-Wierzbowska}, {Goyal}  \&
  {{\.Z}ywucka}}{{Kozie{\l}-Wierzbowska} et~al.}{2020}]{KW2020}
{Kozie{\l}-Wierzbowska} D.,  {Goyal} A.,   {{\.Z}ywucka} N.,  2020, \mn@doi
  [\apjs] {10.3847/1538-4365/ab63d3}, \href
  {https://ui.adsabs.harvard.edu/abs/2020ApJS..247...53K} {247, 53}

\bibitem[\protect\citeauthoryear{{Krizhevsky}, {Sutskever}  \&
  {Hinton}}{{Krizhevsky} et~al.}{2012}]{Krizhevsky2012}
{Krizhevsky} A.,  {Sutskever} I.,   {Hinton} G.~E.,  2012. Advances in Neural
  Information Processing Systems.
ImageNet Classification with Deep Convolutional Neural Networks, Lake Tahoe,
  Nevada, p.~1097

\bibitem[\protect\citeauthoryear{{Lacy} et~al.,}{{Lacy}
  et~al.}{2020}]{Lacy2020}
{Lacy} M.,  et~al., 2020, \mn@doi [\pasp] {10.1088/1538-3873/ab63eb}, \href
  {https://ui.adsabs.harvard.edu/abs/2020PASP..132c5001L} {132, 035001}

\bibitem[\protect\citeauthoryear{{Laing}, {Jenkins}, {Wall}  \&
  {Unger}}{{Laing} et~al.}{1994}]{Laing1994}
{Laing} R.~A.,  {Jenkins} C.~R.,  {Wall} J.~V.,   {Unger} S.~W.,  1994, in
  {Bicknell} G.~V.,  {Dopita} M.~A.,   {Quinn} P.~J.,  eds,  ASP Conf. Ser.
  Vol. 54, The First Stromlo Symposium: The Physics of Active Galaxies. Astron.
  Soc. Pac., San Francisco, p.~201

\bibitem[\protect\citeauthoryear{LeCun, Bottou, Bengio  \& Haffner}{LeCun
  et~al.}{1998}]{LeCun1998}
LeCun Y.,  Bottou L.,  Bengio Y.,   Haffner P.,  1998, Proc. of the IEEE, 86,
  2278

\bibitem[\protect\citeauthoryear{{Leahy} \& {Williams}}{{Leahy} \&
  {Williams}}{1984}]{Leahy1984}
{Leahy} J.~P.,  {Williams} A.~G.,  1984, \mn@doi [\mnras]
  {10.1093/mnras/210.4.929}, \href
  {https://ui.adsabs.harvard.edu/abs/1984MNRAS.210..929L} {210, 929}

\bibitem[\protect\citeauthoryear{Lintott et~al.,}{Lintott
  et~al.}{2008}]{Lintott2008}
Lintott C.~J.,  et~al., 2008, \mn@doi [\mnras]
  {10.1111/j.1365-2966.2008.13689.x}, 389, 1179

\bibitem[\protect\citeauthoryear{{Lukic}, {Br{\"u}ggen}, {Banfield}, {Wong},
  {Rudnick}, {Norris}  \& {Simmons}}{{Lukic} et~al.}{2018}]{Lukic2018}
{Lukic} V.,  {Br{\"u}ggen} M.,  {Banfield} J.~K.,  {Wong} O.~I.,  {Rudnick} L.,
   {Norris} R.~P.,   {Simmons} B.,  2018, \mn@doi [\mnras]
  {10.1093/mnras/sty163}, \href
  {https://ui.adsabs.harvard.edu/abs/2018MNRAS.476..246L} {476, 246}

\bibitem[\protect\citeauthoryear{{Lukic}, {de Gasperin}  \&
  {Br{\"u}ggen}}{{Lukic} et~al.}{2019a}]{Lukic2019b}
{Lukic} V.,  {de Gasperin} F.,   {Br{\"u}ggen} M.,  2019a, \mn@doi [Galaxies]
  {10.3390/galaxies8010003}, \href
  {https://ui.adsabs.harvard.edu/abs/2019Galax...8....3L} {8, 3}

\bibitem[\protect\citeauthoryear{{Lukic}, {Br{\"u}ggen}, {Mingo}, {Croston},
  {Kasieczka}  \& {Best}}{{Lukic} et~al.}{2019b}]{Lukic2019a}
{Lukic} V.,  {Br{\"u}ggen} M.,  {Mingo} B.,  {Croston} J.~H.,  {Kasieczka} G.,
   {Best} P.~N.,  2019b, \mn@doi [\mnras] {10.1093/mnras/stz1289}, \href
  {https://ui.adsabs.harvard.edu/abs/2019MNRAS.487.1729L} {487, 1729}

\bibitem[\protect\citeauthoryear{Luo, Li, Urtasun  \& Zemel}{Luo
  et~al.}{2016}]{luo2016}
Luo W.,  Li Y.,  Urtasun R.,   Zemel R.,  2016. Advances in Neural Information
  Processing Systems.
Understanding the effective receptive field in deep Convolutional Neural
  Networks, Barcelona, Spain, p.~4898

\bibitem[\protect\citeauthoryear{Ma et~al.,}{Ma et~al.}{2019}]{Ma2019}
Ma Z.,  et~al., 2019, \mn@doi [\apjs] {10.3847/1538-4365/aaf9a2}, \href
  {http://dx.doi.org/10.3847/1538-4365/aaf9a2} {240, 34}

\bibitem[\protect\citeauthoryear{Marcos, Volpi  \& Tuia}{Marcos
  et~al.}{2016}]{marcos2016}
Marcos D.,  Volpi M.,   Tuia D.,  2016. 23rd International Conference on
  Pattern Recognition.
Learning rotation invariant convolutional filters for texture classification,
  Cancun, Mexico, p.~2012

\bibitem[\protect\citeauthoryear{Markoff}{Markoff}{2012}]{markoff_2012}
Markoff J.,  2012, Seeking a Better Way to Find Web Images, \url
  {https://www.nytimes.com/2012/11/20/science/for-web-images-creating-new-technology-to-seek-and-find.html}

\bibitem[\protect\citeauthoryear{McGlynn, Scollick  \& White}{McGlynn
  et~al.}{1998}]{mcglynn1998}
McGlynn T.,  Scollick K.,   White N.,  1998, in McLean B.~J.,  A G.~D.,  Hayes
  J.~J.,   Payne H.~E.,  eds,  IAU Symp. Vol. 179, New Horizons from
  Multi-Wavelength Sky Surveys. Kluwer, Dordrecht, p.~465

\bibitem[\protect\citeauthoryear{{Mingo} et~al.,}{{Mingo}
  et~al.}{2019}]{Mingo2019}
{Mingo} B.,  et~al., 2019, \mn@doi [\mnras] {10.1093/mnras/stz1901}, \href
  {https://ui.adsabs.harvard.edu/abs/2019MNRAS.488.2701M} {488, 2701}

\bibitem[\protect\citeauthoryear{{Miraghaei} \& {Best}}{{Miraghaei} \&
  {Best}}{2017}]{Miraghaei2017}
{Miraghaei} H.,  {Best} P.~N.,  2017, \mn@doi [\mnras] {10.1093/mnras/stx007},
  \href {https://ui.adsabs.harvard.edu/abs/2017MNRAS.466.4346M} {466, 4346}

\bibitem[\protect\citeauthoryear{{Missaglia}, {Massaro}, {Capetti}, {Paolillo},
  {Kraft}, {Baldi}  \& {Paggi}}{{Missaglia} et~al.}{2019}]{Missaglia2019}
{Missaglia} V.,  {Massaro} F.,  {Capetti} A.,  {Paolillo} M.,  {Kraft} R.~P.,
  {Baldi} R.~D.,   {Paggi} A.,  2019, \mn@doi [\aap]
  {10.1051/0004-6361/201935058}, \href
  {https://ui.adsabs.harvard.edu/abs/2019A&A...626A...8M} {626, A8}

\bibitem[\protect\citeauthoryear{{Muhammed}, {Ahmed}  \& {Khalid}}{{Muhammed}
  et~al.}{2017}]{Muhammed2017}
{Muhammed} M. A.~E.,  {Ahmed} A.~A.,   {Khalid} T.~A.,  2017, in 2017
  International Conference On Smart Technologies For Smart Nation
  (SmartTechCon). pp 902--907, \mn@doi{10.1109/SmartTechCon.2017.8358502}

\bibitem[\protect\citeauthoryear{Neelakantan, Vilnis, Le, Sutskever, Kaiser,
  Kurach  \& Martens}{Neelakantan et~al.}{2015}]{neelakantan2015}
Neelakantan A.,  Vilnis L.,  Le Q.~V.,  Sutskever I.,  Kaiser L.,  Kurach K.,
  Martens J.,  2015, preprint (arXiv:1511.06807)

\bibitem[\protect\citeauthoryear{{Norris} et~al.,}{{Norris}
  et~al.}{2011}]{Norris2011}
{Norris} R.~P.,  et~al., 2011, \mn@doi [\pasa] {10.1071/AS11021}, \href
  {https://ui.adsabs.harvard.edu/abs/2011PASA...28..215N} {28, 215}

\bibitem[\protect\citeauthoryear{{Norris} et~al.,}{{Norris}
  et~al.}{2013}]{Norris2013}
{Norris} R.~P.,  et~al., 2013, \mn@doi [\pasa] {10.1017/pas.2012.020}, \href
  {https://ui.adsabs.harvard.edu/abs/2013PASA...30...20N} {30, e020}

\bibitem[\protect\citeauthoryear{{Ocran}, {Taylor}, {Vaccari},
  {Ishwara-Chandra}  \& {Prandoni}}{{Ocran} et~al.}{2020}]{Ocran2020}
{Ocran} E.~F.,  {Taylor} A.~R.,  {Vaccari} M.,  {Ishwara-Chandra} C.~H.,
  {Prandoni} I.,  2020, \mn@doi [\mnras] {10.1093/mnras/stz2954}, \href
  {https://ui.adsabs.harvard.edu/abs/2020MNRAS.491.1127O} {491, 1127}

\bibitem[\protect\citeauthoryear{{Owen} \& {Ledlow}}{{Owen} \&
  {Ledlow}}{1994}]{Owen1994}
{Owen} F.~N.,  {Ledlow} M.~J.,  1994, in {Bicknell} G.~V.,  {Dopita} M.~A.,
  {Quinn} P.~J.,  eds,  ASP Conf. Ser. Vol. 54, The First Stromlo Symposium:
  The Physics of Active Galaxies. Astron. Soc. Pac., San Francisco, p.~319

\bibitem[\protect\citeauthoryear{{Owen} \& {Rudnick}}{{Owen} \&
  {Rudnick}}{1976}]{Owen1976}
{Owen} F.~N.,  {Rudnick} L.,  1976, \mn@doi [\apjl] {10.1086/182077}, \href
  {https://ui.adsabs.harvard.edu/abs/1976ApJ...205L...1O} {205, L1}

\bibitem[\protect\citeauthoryear{{Pracy} et~al.,}{{Pracy}
  et~al.}{2016}]{Pracy2016}
{Pracy} M.~B.,  et~al., 2016, \mn@doi [\mnras] {10.1093/mnras/stw910}, \href
  {https://ui.adsabs.harvard.edu/abs/2016MNRAS.460....2P} {460, 2}

\bibitem[\protect\citeauthoryear{{Prescott} et~al.,}{{Prescott}
  et~al.}{2018}]{Prescott2018}
{Prescott} M.,  et~al., 2018, \mn@doi [\mnras] {10.1093/mnras/sty1789}, \href
  {https://ui.adsabs.harvard.edu/abs/2018MNRAS.480..707P} {480, 707}

\bibitem[\protect\citeauthoryear{{Proctor}}{{Proctor}}{2011}]{Proctor2011}
{Proctor} D.~D.,  2011, \mn@doi [\apjs] {10.1088/0067-0049/194/2/31}, \href
  {https://ui.adsabs.harvard.edu/abs/2011ApJS..194...31P} {194, 31}

\bibitem[\protect\citeauthoryear{{Roberts}, {Saripalli}, {Wang},
  {Sathyanarayana Rao}, {Subrahmanyan}, {KleinStern}, {Morii-Sciolla}  \&
  {Simpson}}{{Roberts} et~al.}{2018}]{Roberts2018}
{Roberts} D.~H.,  {Saripalli} L.,  {Wang} K.~X.,  {Sathyanarayana Rao} M.,
  {Subrahmanyan} R.,  {KleinStern} C.~C.,  {Morii-Sciolla} C.~Y.,   {Simpson}
  L.,  2018, \mn@doi [\apj] {10.3847/1538-4357/aa9c49}, \href
  {https://ui.adsabs.harvard.edu/abs/2018ApJ...852...47R} {852, 47}

\bibitem[\protect\citeauthoryear{{Rudnick} \& {Owen}}{{Rudnick} \&
  {Owen}}{1977}]{Rudnick1977}
{Rudnick} L.,  {Owen} F.~N.,  1977, \mn@doi [\aj] {10.1086/112001}, \href
  {https://ui.adsabs.harvard.edu/abs/1977AJ.....82....1R} {82, 1}

\bibitem[\protect\citeauthoryear{Sabour, Frosst  \& Hinton}{Sabour
  et~al.}{2017}]{sabour2017}
Sabour S.,  Frosst N.,   Hinton G.~E.,  2017, in Proceedings of the 31st
  International Conference on Neural Information Processing Systems. p.~3856

\bibitem[\protect\citeauthoryear{{Sadler}, {Ekers}, {Mahony}, {Mauch}  \&
  {Murphy}}{{Sadler} et~al.}{2014}]{Sadler2014}
{Sadler} E.~M.,  {Ekers} R.~D.,  {Mahony} E.~K.,  {Mauch} T.,   {Murphy} T.,
  2014, \mn@doi [\mnras] {10.1093/mnras/stt2239}, \href
  {https://ui.adsabs.harvard.edu/abs/2014MNRAS.438..796S} {438, 796}

\bibitem[\protect\citeauthoryear{{Sadr}, {Vos}, {Bassett}, {Hosenie}, {Oozeer}
  \& {Lochner}}{{Sadr} et~al.}{2019}]{Sadr2019}
{Sadr} A.~V.,  {Vos} E.~E.,  {Bassett} B.~A.,  {Hosenie} Z.,  {Oozeer} N.,
  {Lochner} M.,  2019, \mn@doi [\mnras] {10.1093/mnras/stz131}, \href
  {https://ui.adsabs.harvard.edu/abs/2019MNRAS.484.2793V} {484, 2793}

\bibitem[\protect\citeauthoryear{{Sandage}}{{Sandage}}{1961}]{Sandage1961}
{Sandage} A.,  1961, {The Hubble Atlas of Galaxies}

\bibitem[\protect\citeauthoryear{Simonyan \& Zisserman}{Simonyan \&
  Zisserman}{2015}]{Simonyan2014}
Simonyan K.,  Zisserman A.,  2015, in International Conference on Learning
  Representations.

\bibitem[\protect\citeauthoryear{{Smith} \& {Donohoe}}{{Smith} \&
  {Donohoe}}{2019}]{Smith2019}
{Smith} M.~D.,  {Donohoe} J.,  2019, \mn@doi [\mnras] {10.1093/mnras/stz2525},
  \href {https://ui.adsabs.harvard.edu/abs/2019MNRAS.490.1363S} {490, 1363}

\bibitem[\protect\citeauthoryear{Srivastava, Hinton, Krizhevsky, Sutskever  \&
  Salakhutdinov}{Srivastava et~al.}{2014}]{Srivastava2014}
Srivastava N.,  Hinton G.,  Krizhevsky A.,  Sutskever I.,   Salakhutdinov R.,
  2014, J. Mach. Learn. Res., 15, 1929–1958

\bibitem[\protect\citeauthoryear{{Tang}, {Scaife}  \& {Leahy}}{{Tang}
  et~al.}{2019}]{Tang2019}
{Tang} H.,  {Scaife} A.~M.~M.,   {Leahy} J.~P.,  2019, \mn@doi [\mnras]
  {10.1093/mnras/stz1883}, \href
  {https://ui.adsabs.harvard.edu/abs/2019MNRAS.488.3358T} {488, 3358}

\bibitem[\protect\citeauthoryear{{Whittam}, {Green}, {Jarvis}  \&
  {Riley}}{{Whittam} et~al.}{2020}]{Whittam2020}
{Whittam} I.~H.,  {Green} D.~A.,  {Jarvis} M.~J.,   {Riley} J.~M.,  2020,
  \mn@doi [\mnras] {10.1093/mnras/staa306}, \href
  {https://ui.adsabs.harvard.edu/abs/2020MNRAS.493.2841W} {493, 2841}

\bibitem[\protect\citeauthoryear{{Willett} et~al.,}{{Willett}
  et~al.}{2013}]{Willett2013}
{Willett} K.~W.,  et~al., 2013, \mn@doi [\mnras] {10.1093/mnras/stt1458}, \href
  {https://ui.adsabs.harvard.edu/abs/2013MNRAS.435.2835W} {435, 2835}

\bibitem[\protect\citeauthoryear{{Wu} et~al.,}{{Wu} et~al.}{2019}]{Wu2019}
{Wu} C.,  et~al., 2019, \mn@doi [\mnras] {10.1093/mnras/sty2646}, \href
  {https://ui.adsabs.harvard.edu/abs/2019MNRAS.482.1211W} {482, 1211}

\bibitem[\protect\citeauthoryear{{de Vaucouleurs}}{{de
  Vaucouleurs}}{1959}]{deVaucouleurs1959}
{de Vaucouleurs} G.,  1959, \mn@doi [Handbuch der Physik]
  {10.1007/978-3-642-45932-0_7}, \href
  {https://ui.adsabs.harvard.edu/abs/1959HDP....53..275D} {53, 275}

\makeatother
\end{thebibliography}
%



\begin{landscape}

\begin{table}

\appendix
\section{Table of Results}

\caption{Results from the experiments discussed in Section~\ref{sec:experiments}, each architecture's name and the keys given in Table~\ref{tab:arch_keys} for use in the figures. $\dagger$ denotes that the architecture have been modified due to computational constraints or re-purposed for classification. $\ddagger$ the results given for the inference time and images classified per second are for the classification of 3072 images at batch size 32.}
\label{tab:table_results_1}
\setlength{\tabcolsep}{2.9pt}
\begin{tabular}{llrrrrllrlllr}
\hline
Architecture Name        & Key  & Floating Point & Convolutional & Fully Connected & Trainable & Inference Time$\ddagger$ & St. Dev. & Images Classed & Effective & Effective  & Effective & GPU Memory \\
 &   & Operations (FLOPs) & FLOPs & FLOPs & Parameters & (Seconds) & (Seconds) & per Second$\ddagger$ & Receptive Field & Stride & Padding & Usage (MB) \\
\hline

{\sc AlexNet}                  & ALN  & 1107736302                        & 1074169584          & 33566718              & 37302980             & 2.72816                  & 0.24097                      & 1126              & 195                       & 32               & 0                 & 280.576               \\
{\sc ATLAS X-ID}$\dagger$      & ATL  & 546585022                         & 545304800           & 1280222               & 1385988              & 2.90696                  & 0.36538                      & 1056              & 64                        & 10               & 24                & 411.648               \\
{\sc CLARAN} ({\sc VGG16D})$\dagger$ & VGG  & 27231525758                       & 27044866944         & 186658814             & 201384644            & 10.52742                 & 0.33702                      & 291               & 212                       & 32               & 90                & 4061.184              \\
{\sc ConvNet4}                 & CN4  & 1036529876                        & 870640128           & 165889748             & 165910168            & 3.51457                  & 0.22828                      & 874               & 24                        & 4                & 12                & 897.024               \\
{\sc ConvNet8}                 & CN8  & 4612528724                        & 4571054976          & 41473748              & 42646184             & 4.73292                  & 0.24774                      & 649               & 76                        & 16               & 30                & 1844.224              \\
{\sc ConvXpress}               & CXP  & 764997460                         & 762947712           & 2049748               & 3415944              & 2.9279                   & 0.21564                      & 1049              & 333                       & 64               & 136               & 393.216               \\
{\sc FIRST Classifier}         & 1stC & 1128841236                        & 1077316875          & 51524361              & 51655412             & 3.29737                  & 0.26081                      & 931               & 22                        & 8                & 7                 & 1715.2                \\
{\sc FR-Deep}                  & FR-D & 141486958                         & 141023472           & 463486                & 479996               & 2.88221                  & 0.29405                      & 1065              & 84                        & 30               & 27                & 412.672               \\
{\sc Hosenie}                  & H    & 109649469                         & 109413751           & 235718                & 261239               & 2.76616                  & 0.2965                       & 1110              & 94                        & 18               & 38                & 315.392               \\
{\sc MCRGNet}$\dagger$         & MCRG & 8406674                           & 8201652             & 205022                & 213916               & 2.41828                  & 0.27093                      & 1270              & 63                        & 32               & 20                & 63.488                \\
{\sc Radio Galaxy Zoo}         & RGZ  & 75825974                          & 70580024            & 5245950               & 5283444              & 2.46538                  & 0.31239                      & 1246              & 74                        & 26               & 8                 & 84.992                \\
{\sc SimpleNet}                & CNs  & 797660134                         & 797639400           & 20734                 & 37460                & 3.69593                  & 0.2417                       & 831               & 49                        & 16               & 11                & 573.44                \\
{\sc Toothless}$\dagger$       & TLS  & 1634645742                        & 1566476016          & 68169726              & 71906180             & 2.94229                  & 0.21572                      & 1044              & 195                       & 32               & 64                & 782.336              
   
\\
\hline
\end{tabular}
\end{table}

\begin{table}
\setcounter{table}{1}
\renewcommand{\thetable}{A\arabic{table}}
\caption{Results of from the recognition performance experiments discussed in Section~\ref{sec:experiments}, each architecture's name and the keys given in Table~\ref{tab:arch_keys} for use in the figures.}
\label{tab:table_results_2}
\begin{tabular}{lllllllllllllll}
\hline
Architecture Name & Key & Mean per Class & Precision & & & & Recall & & & & F1-Score & & &  \\
 & & Accuracy & (Compact) & (FRI) & (FRII) & (Bent) & (Compact) & (FRI) & (FRII) & (Bent) & (Compact) & (FRI) & (FRII) & (Bent) \\
\hline
{\sc AlexNet}                  & ALN  & 63.15                   & 0.837               & 0.7837          & 0.5022           & 0.2509           & 0.8634           & 0.5872       & 0.6785        & 0.397         & 0.8488             & 0.6698         & 0.5734          & 0.3007          \\
{\sc ATLAS X-ID}$\dagger$      & ATL  & 65.34                   & 0.8782              & 0.7784          & 0.5636           & 0.2069           & 0.853            & 0.6491       & 0.5341        & 0.5776        & 0.8639             & 0.7047         & 0.5461          & 0.3043          \\
{\sc CLARAN} ({\sc VGG16D})$\dagger$ & VGG  & 72.98                   & 0.893               & 0.8225          & 0.636            & 0.2658           & 0.8661           & 0.6621       & 0.6775        & 0.7137        & 0.879              & 0.729          & 0.6531          & 0.3861          \\
{\sc ConvNet4}                 & CN4  & 62.84                   & 0.8678              & 0.7586          & 0.5639           & 0.2025           & 0.8401           & 0.6662       & 0.6336        & 0.3735        & 0.852              & 0.707          & 0.5913          & 0.2576          \\
{\sc ConvNet8}                 & CN8  & 71.71                   & 0.8925              & 0.7923          & 0.6665           & 0.3065           & 0.8687           & 0.7242       & 0.644         & 0.6314        & 0.8798             & 0.756          & 0.6475          & 0.4126          \\
{\sc ConvXpress}               & CXP  & 71.75                   & 0.8983              & 0.7849          & 0.6527           & 0.2878           & 0.858            & 0.7065       & 0.6405        & 0.6648        & 0.8767             & 0.7433         & 0.6405          & 0.4017          \\
{\sc FIRST Classifier}         & 1stC & 65.21                   & 0.8697              & 0.7408          & 0.585            & 0.2698           & 0.8279           & 0.6937       & 0.6341        & 0.4527        & 0.8479             & 0.713          & 0.6056          & 0.3348          \\
{\sc FR-Deep}                  & FR-D & 59.36                   & 0.866               & 0.7451          & 0.5282           & 0.1552           & 0.8433           & 0.6243       & 0.2199        & 0.6871        & 0.8532             & 0.6773         & 0.2956          & 0.2527          \\
{\sc Hosenie}                  & H    & 58.5                    & 0.8957              & 0.7014          & 0.3894           & 0.1755           & 0.8216           & 0.6837       & 0.2238        & 0.611         & 0.857              & 0.6923         & 0.2809          & 0.2717          \\
{\sc MCRGNet}$\dagger$         & MCRG & 68.09                   & 0.8686              & 0.7809          & 0.6367           & 0.2281           & 0.8658           & 0.6499       & 0.59          & 0.6178        & 0.867              & 0.709          & 0.6124          & 0.333           \\
{\sc Radio Galaxy Zoo}         & RGZ  & 69.87                   & 0.886               & 0.8102          & 0.633            & 0.2144           & 0.8662           & 0.6292       & 0.6425        & 0.6568        & 0.876              & 0.7076         & 0.633           & 0.3229          \\
{\sc SimpleNet}                & CNs  & 65.91                   & 0.8763              & 0.7776          & 0.5758           & 0.2096           & 0.8593           & 0.6446       & 0.5563        & 0.5764        & 0.8674             & 0.7037         & 0.5657          & 0.3065          \\
{\sc Toothless}$\dagger$       & TLS  & 68.92                   & 0.8973              & 0.785           & 0.614            & 0.2083           & 0.8189           & 0.6542       & 0.6518        & 0.632         & 0.8548             & 0.7085         & 0.6315          & 0.3129         

 \\       

\hline
\end{tabular}
\end{table}
\bsp	
\label{lastpage}
\end{landscape}

\end{document}